\documentclass{ws-jai}
\usepackage[flushleft]{threeparttable}
\begin{document}

\catchline{}{}{}{}{} 

\markboth{Risacher}{The upGREAT dual frequency heterodyne arrays for SOFIA}

\title{The upGREAT dual frequency heterodyne arrays for SOFIA}

\author{C. Risacher$^{1,6*}$, R. G\"usten$^{1}$, J. Stutzki$^{2}$, H.-W. H\"ubers$^{3}$, 
  R. Aladro$^{1}$, A. Bell$^{1}$, C. Buchbender$^{2}$,  D. B\"uchel$^{2}$, T. Csengeri$^{1}$, C. Duran$^{1}$, 
 U. U. Graf$^{2}$,  R. D. Higgins$^{2}$, C. E. Honingh$^{2}$, K. Jacobs$^{2}$, M. Justen$^{2}$, 
  B. Klein$^{1,5}$ , M. Mertens$^{2}$, Y. Okada$^{2}$, A. Parikka$^{2}$, P. P\"utz$^{2}$, N. Reyes$^{1,4}$, H. Richter$^{3}$, O. Ricken$^{1}$, 
  D. Riquelme$^{1}$, N. Rothbart$^{3}$, N. Schneider$^{2}$, R. Simon$^{2}$, M. Wienold$^{3}$, H. Wiesemeyer$^{1}$, M. Ziebart$^{2}$, P. Fusco$^{7}$, S. Rosner$^{7, 8}$ \and B. Wohler$^{7, 8}$ 
}

\address{
$^{1}$Max-Planck-Institut f\"ur Radioastronomie, Auf dem H\"ugel 69, 53121, Bonn, Germany, crisache@mpifr.de\\
$^{2}$I. Physikalisches Institut der Universit\"at zu K\"oln, Z\"ulpicher Strasse 77, 50937 K\"oln, Germany\\
$^{3}$Institute of Optical Sensor Systems, German Aerospace Center (DLR), Rutherfordstr. 2, 12489 Berlin, Germany\\
$^{4}$Departamento de Ingenier\'{i}a El\'ectrica, Universidad de Chile, Santiago, Chile\\
$^{5}$University of Applied Sciences Bonn-Rhein-Sieg, Sankt Augustin, 53757 Germany\\
$^{6}$IRAM, 300 rue de la Piscine, 38406 Saint Martin d'Heres, France\\
$^{7}$NASA Ames Research Center, Moffett Field, CA 94035, USA \\
$^{8}$SETI Institute, Mountain View, CA 94043, USA 
}

\maketitle

\corres{$^{*}$Corresponding author.}

\begin{history}
\received{(to be inserted by publisher)};
\revised{(to be inserted by publisher)};
\accepted{(to be inserted by publisher)};
\end{history}

\begin{abstract}
We present the performance of the upGREAT heterodyne array receivers on the SOFIA telescope after several years of operations. This instrument is a multi-pixel high resolution (R $\gtrsim$ $10^7$) spectrometer for the Stratospheric Observatory for Far-Infrared Astronomy (SOFIA). The receivers use 7-pixel subarrays configured in a hexagonal layout around a central pixel. The low frequency array receiver (LFA) has 2x7 pixels (dual polarization), and presently covers the 1.83-2.06 THz frequency range, which allows to observe the [CII]  and [OI] lines at 158 ${\mu}m$ and 145 ${\mu}m$ wavelengths. The high frequency array (HFA) covers the [OI] line at 63 ${\mu}m$ and is equipped with one polarization at the moment (7 pixels, which can be upgraded in the near future with a second polarization array). The 4.7 THz array has successfully flown using two separate quantum-cascade laser local oscillators from two different groups. NASA completed the development, integration and testing of a dual-channel closed-cycle cryocooler system, with two independently operable He compressors, aboard SOFIA in early 2017 and since then, both arrays can be operated in parallel using a frequency separating dichroic mirror. This configuration is now the prime GREAT configuration and has been added to SOFIA's instrument suite since observing cycle 6.  
\end{abstract}


\section{Introduction - Scientific motivation}
\noindent 
 
Since the turn-off of the HIFI/Herschel instrument \cite{degrauuw2010}, there has been very limited number of high-resolution spectroscopy instruments capable of performing far-infrared wavelengths observations. For example, the balloon-borne Stratospheric Terahertz Observatory (STO-2) performed high-resolution observations in the THz range in December 2016 for about $\sim$500 hours. On the other hand, the spectrometer GREAT  \cite{heyminck2012} aboard the SOFIA airborne observatory \cite{young2012} is the only facility that provides regular high-resolution spectroscopy at far-infrared wavelengths. 

GREAT saw first light on April 1st 2011 during SOFIA’s early science phase, and since then has been operated on more than 135 science flights, collecting unique science data for more than 1000 hours. Starting as a dual-color single-pixel receiver operating in parallel in the 1.5 and 1.9 THz frequency bands, a major upgrade occurred in 2015 when the low-frequency array receiver upGREAT/LFA was installed, spatially multiplexing now with 2x7 pixels in the 1.9 THz band \cite{risacher2016a}. A year later, in May 2016, the second frequency band, the high-frequency array upGREAT/HFA operating at 4.7 THz was successfully commissioned. As the SOFIA project had only one flight compressor in 2015/2016, only one of the arrays could be operated at a given time. In 2017, the phase 2 SOFIA infrastructure was completed, for the first time allowing two compressors, and therefore two closed cycle systems to operate independently and simultaneously. upGREAT in its full dual-array configuration was then commissioned in May /June 2017 and has been operated since during 25 science flights. This paper describes the development of the instrument and its in-flight performance.

\section{SOFIA observations - atmospheric conditions}

\begin{figure}[h]
\begin{center}
\includegraphics[angle=0,width=4in]{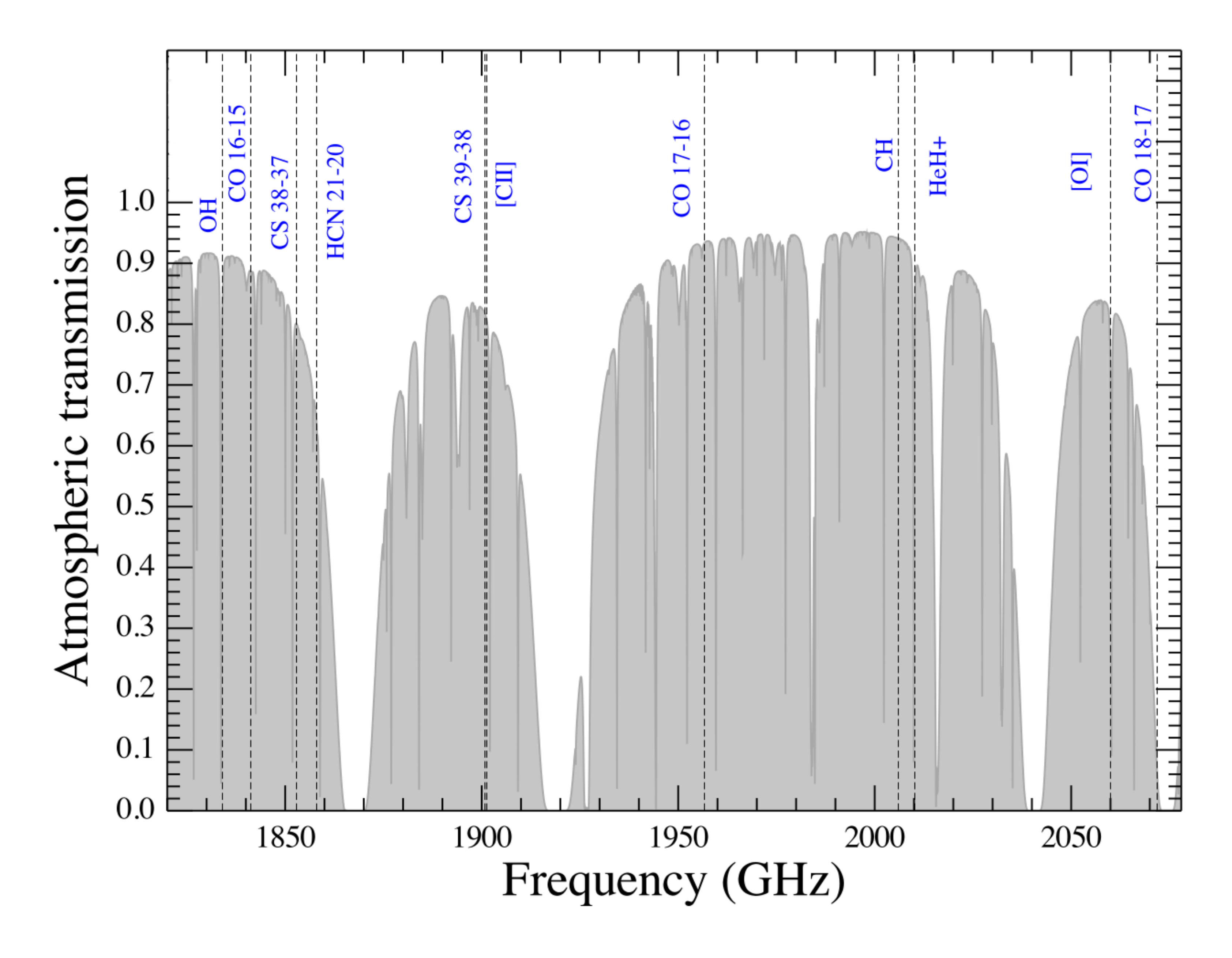} 
\end{center}
\caption{Atmospheric transmission at 43000 feet with 15 ${\mu}m$ water vapor}
\label{atm1}
\end{figure}

\begin{figure}[h]
\begin{center}
\includegraphics[angle=0,width=4in]{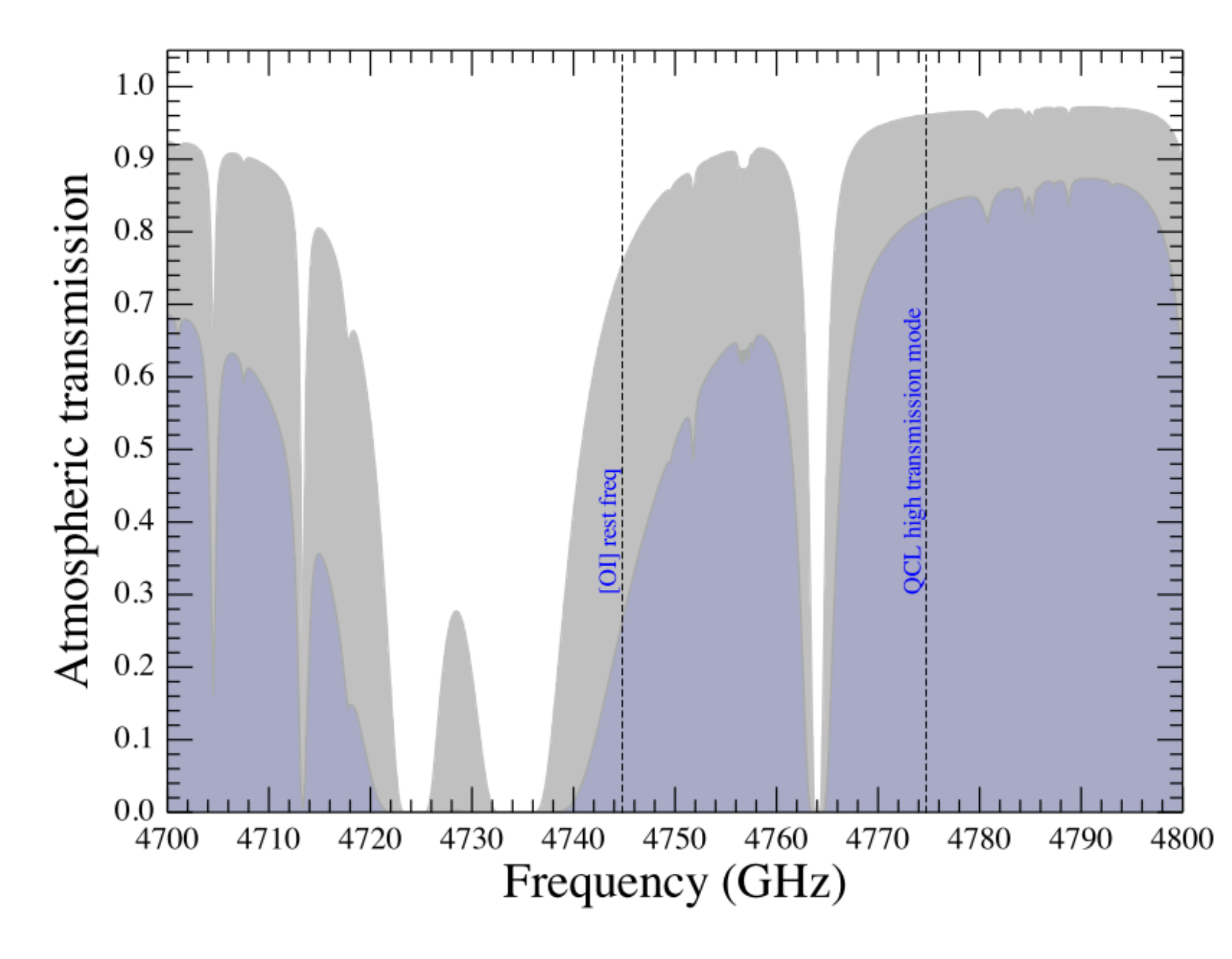} 
\end{center}
\caption{Atmospheric transmission at 43000 feet with 15 ${\mu}m$ water vapor (purple) and 5 ${\mu}m$ (grey). At the [OI] rest frequency, the atmospheric transmission more than doubles in this case}
\label{atm2}
\end{figure}

Observations in the THz range are almost impossible to achieve from ground, therefore only airborne, balloon borne or space telescopes can perform such observations.  For SOFIA, flying regularly at altitudes above 41000 feet, a large part of this spectrum becomes accessible with high atmospheric transmission.  Even though it flies above 99 percent of the Earth's atmosphere, it still suffers from absorption from several atmospheric molecular species, e.g. H${_2}$O, O${_3}$.  Figures \ref{atm1} and \ref{atm2} show the typical atmospheric transmission for the two upGREAT bands, for 1.8-2.07 THz and around 4.745 THz.   Some of the main lines of interest are shown overlaid. Figure \ref{atm1} shows that a large part of the band is not accessible, and even in the regions having good atmospheric transmission, not all frequencies are tunable with the current local oscillators (Sect 4.2). For the higher frequencies, as can be seen in Fig. \ref{atm2}, the [OI] observations are very challenging to perform, even from SOFIA, and require precipitable water vapours (PWV) well below 10 ${\mu}m$, in order not to be too degraded by the atmospheric losses. Such PWV values are difficult to achieve when SOFIA is stationed in Palmdale, CA, where it normally operates. However, SOFIA is deployed yearly to New Zealand and is stationed in Christchurch during the Southern winter.  There, the atmospheric conditions are usually of the order of a few microns of PWV, hence allowing more efficient [OI] observations.

\section{Instrument general description}

A full detailed description of the receivers hardware can be found in \citet{risacher2016a} and the first commissioning results of the low frequency upGREAT receiver are presented in \citet{risacher2016b}. We present here the instrument performance as of 2018, highlighting the performance of the high frequency array for 4.7 THz operation, and describing the improvements and experience gained since the installation of those systems.  Since May 2017, the two upGREAT arrays are operating in parallel, with 21 pixels. The main instrument characteristics are summarised in Table 1.  

\begin{wstable}[h]
\caption{upGREAT receiver characteristics}
\begin{tabular}{@{}cccc@{}} \toprule
 &  LFA &  HFA & Comment\\
 \botrule
RF Bandwidth  & 1.83-2.006 THz * & 4.745 THz $\pm$ few GHz & *up to 2.07 THz (experimental)\\
IF Bandwidth & 0.5-4 GHz & 0.5-4.0 GHz &  \\
Spectral resolution &  $\gtrsim$ $10^7$ &  $\gtrsim$ $10^7$ & \\
Mixer techonology & NbN HEB & NbN HEB & waveguide based \\
LO technology & solid state multipliers & quantum-cascade lasers \\
Receiver Sensitivity & 2000 K & 2500 K & SSB, at 2 GHz IF \\
System noise temperature & 2000-2500 K& 3500 K & weather dependent \\ 
Number of pixels & 7 per sub-array (14 total)  & 7 in one sub-array &  \\
Backends & 4 GHz instantaneous BW & 4 GHz instantaneous BW & FFTS technology \\
Array Geometry & hexagonal with central pixel & hexagonal with central pixel \\
\botrule
\end{tabular}
\label{aba:tbl1}
\end{wstable}

\begin{figure}[h]
\begin{center}
\includegraphics[angle=0,width=6in]{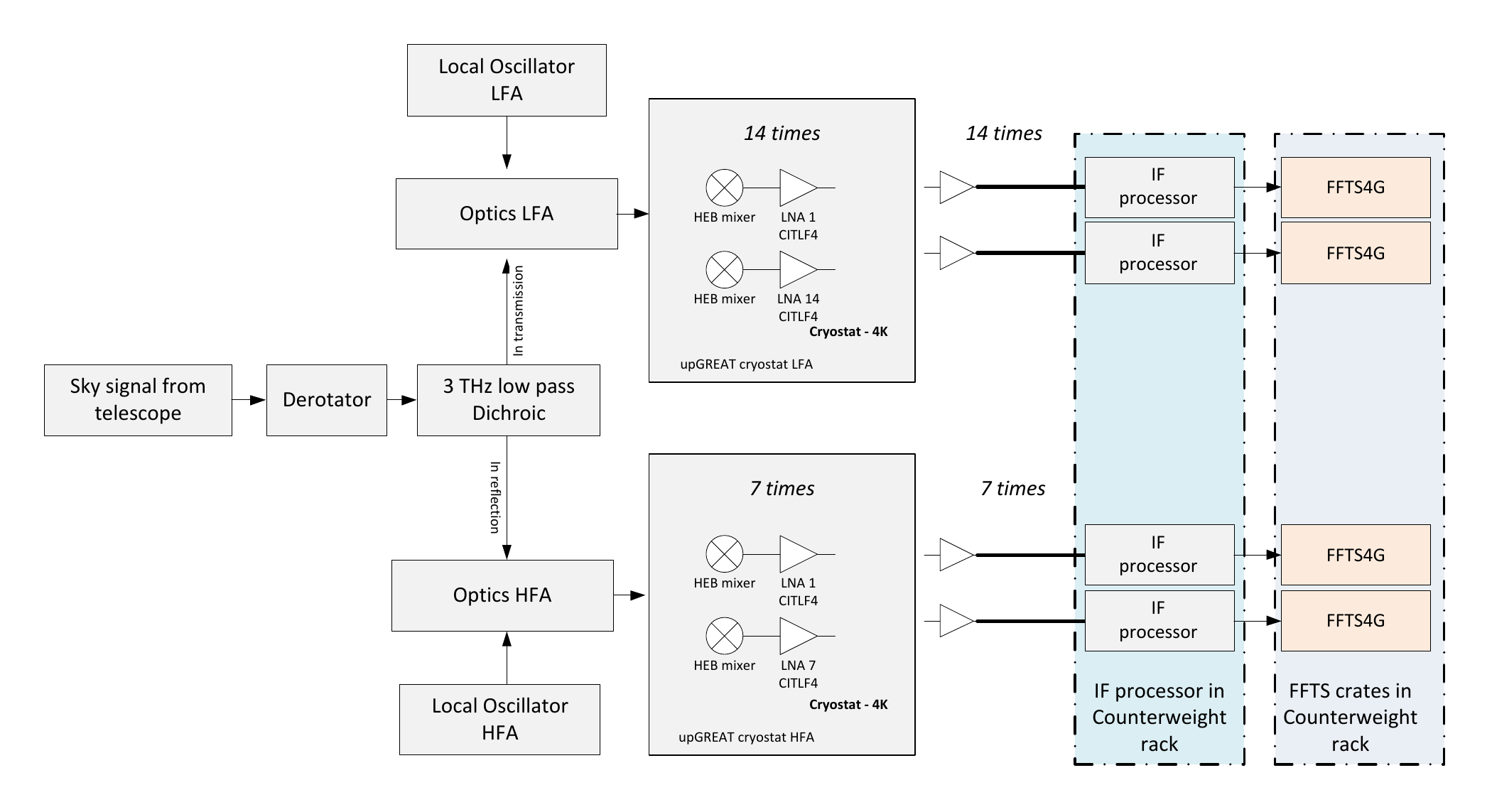} 
\end{center}
\caption{Diagram showing the main components of the upGREAT receiver system - adapted from \citet{risacher2016a}}
\label{schema}
\end{figure}

The main components of the upGREAT LFA/HFA configuration, are shown in Fig. \ref{schema}.  The astronomical signal is directed through the SOFIA telescope main optics and the first optical component from the GREAT system is the derotator optics, a "K-mirror" rotating set of mirrors, which allows to compensate for the sky rotation in the instrument focal plane (described in detail in \citet{risacher2016b}).
The frequency selection is done via a dichroic filter from QMC Instruments Ltd., which separates in transmission the low frequencies below 3 THz (to the upGREAT LFA system), and reflects the 4.7 THz signal (to the HFA receiver) with less than 7\%  and 2\% losses respectively.  This component allows the simultaneous operation of both upGREAT arrays.  Then the Local Oscillator (LO) references are combined with the astronomical signal, via dedicated pre-adjusted optics, which also reimage as needed the optical parameters via pairs of Gaussian telescope optics (details in \citet{risacher2016a} and in following sections). The LO references are located inside elongated housings mounted under the Science Instrument. For the upGREAT/LFA system, the LO systems are based on multiplier chains from VDI Inc. and for the HFA system, the LO system is using quantum-cascade devices (detailed in 5.2 and 5.3). Details on the cryostat, mixer design, IF system and backends can be found in \citet{risacher2016a}.  The superconducting Hot Electron Bolometers (HEB) mixers are developed and fabricated by the I. Physikalisches Institut der Universit{\"a}t zu K{\"o}ln \cite{puetz2012, buechel2015}.

\begin{figure}[h]
\begin{center}
\includegraphics[angle=0,width=6in]{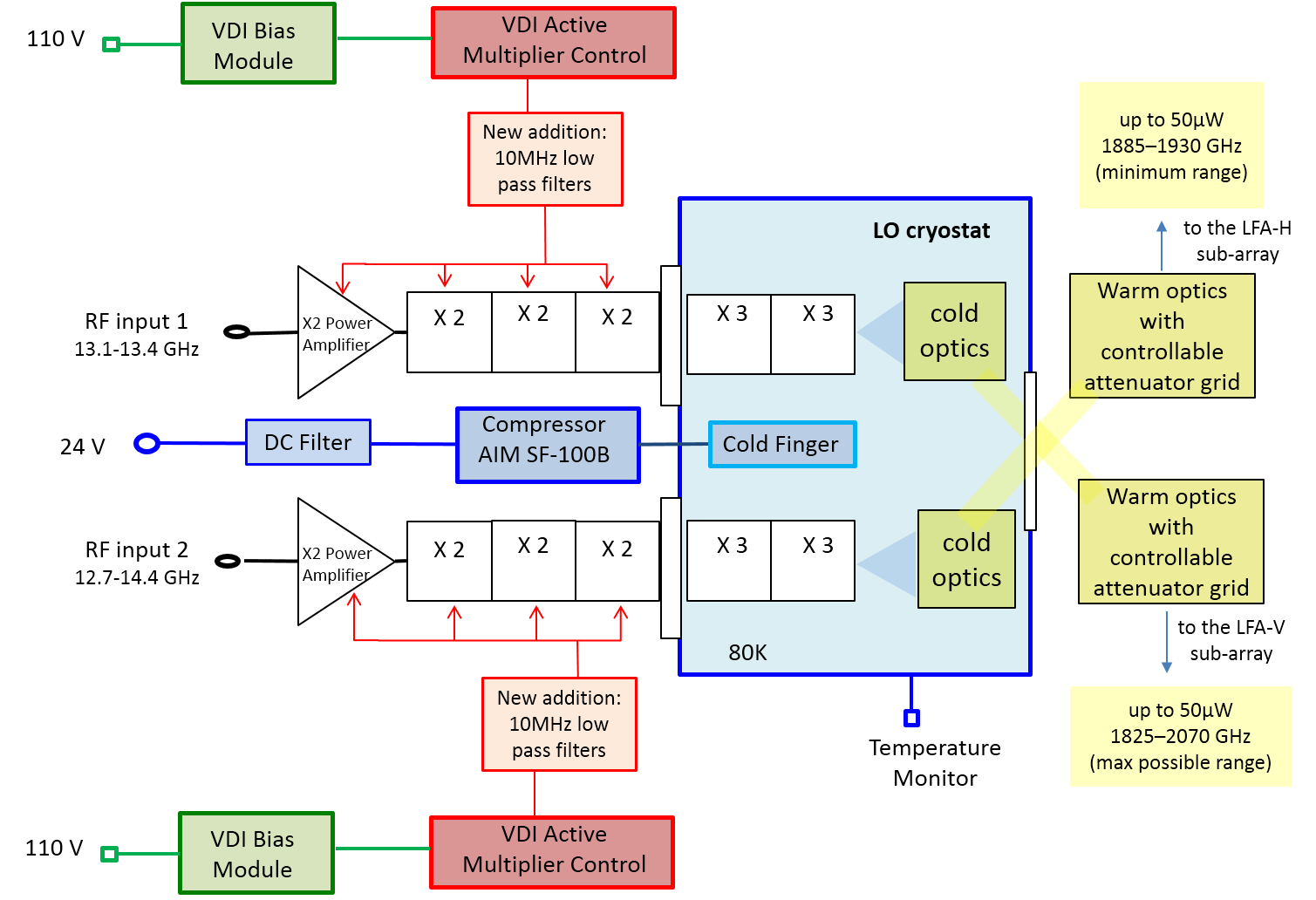} 
\end{center}
\caption{Main LO diagram showing the modifications compared to the original configuration (Fig. 3 in \cite{risacher2016a}). Each LO chain is now driven by independent synthesizers, the LO paths are now completely independent from each other and  the output LO power adjustment is now performed by remotely controllable wire grids. Note also the inclusion of 10 MHz low pass filters in every multiplier bias lines, allowing to avoid spurious lines in the IF bandpass of the mixers.}
\label{LFA_LO_diagram}
\end{figure}

\begin{figure}[h]
\begin{center}
\includegraphics[angle=0,width=4in]{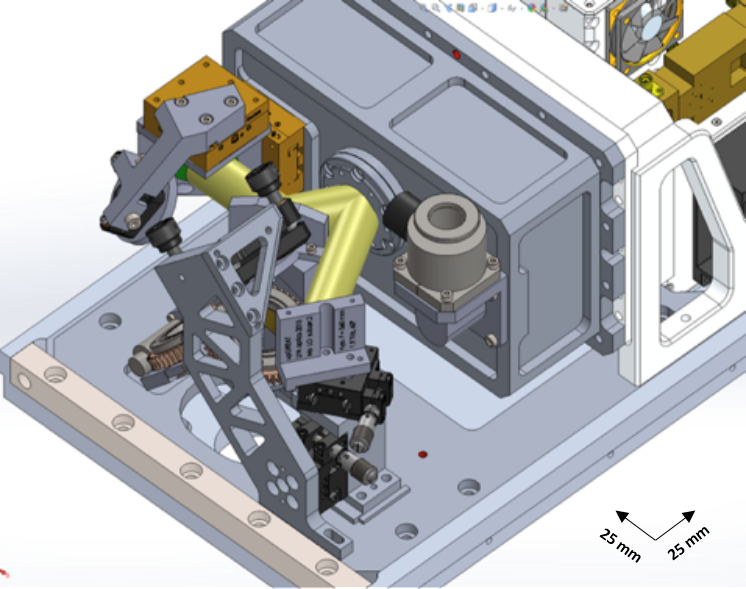} 
\end{center}
\caption{Model showing the updated LO optics. The two LO chains are now fully independent, and their output paths are shown in yellow. Additional wire grids on motorized supports were added in February 2018, allowing to control the LO output levels of both polarizations independently.}
\label{LO1}
\end{figure}

\begin{figure}[h]
\begin{center}
\includegraphics[angle=-90,width=6in]{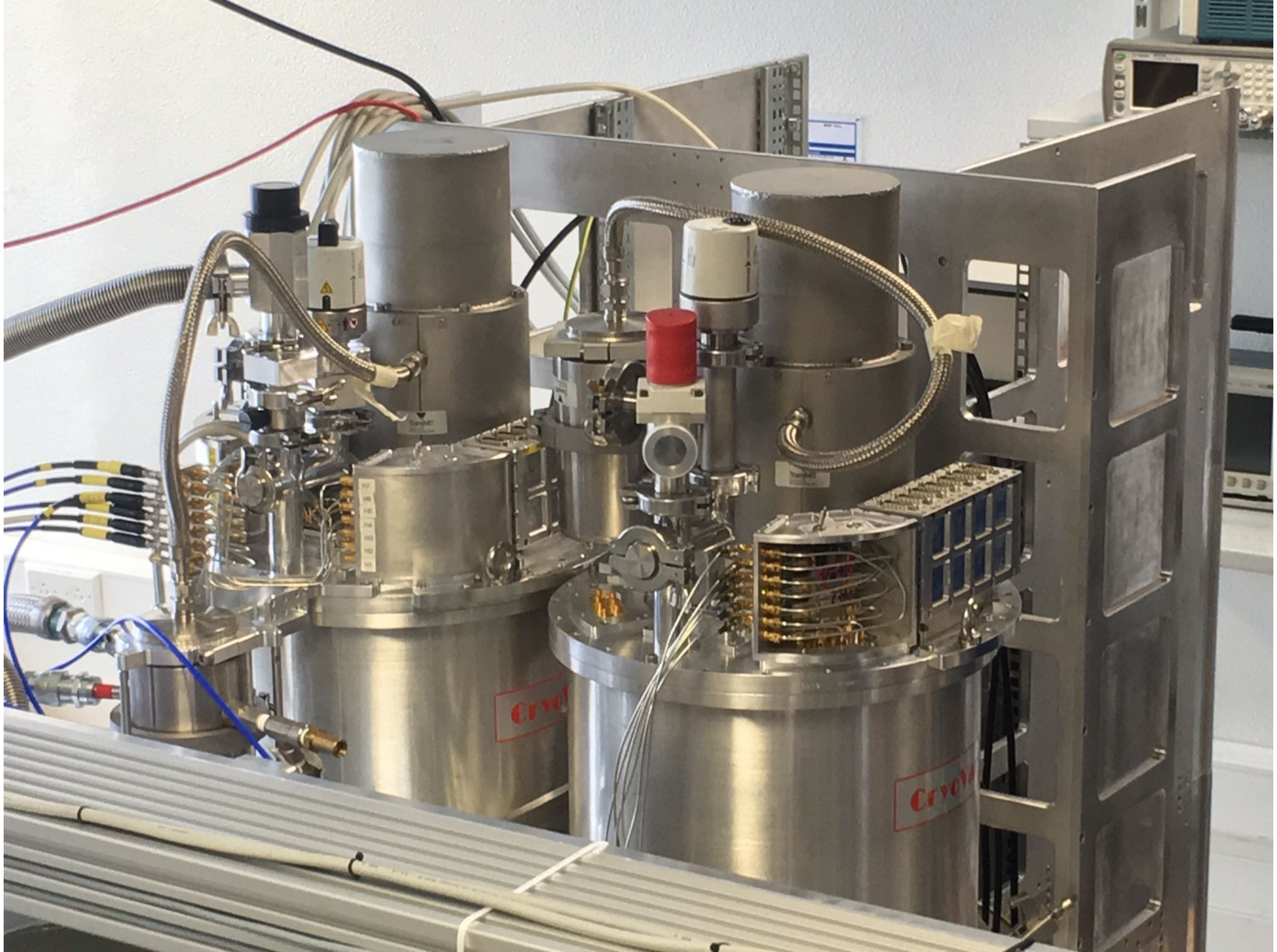} 
\end{center}
\caption{View of the two closed-cycle receivers upGREAT LFA (left) and HFA (right)}
\label{main1}
\end{figure}

\section{LFA array}

\subsection{Initial design and optics modifications}

Compared to the original design for the upGREAT/LFA array, several things were modified since the first installation onboard SOFIA \cite{risacher2016a}. The optics plate and the local oscillator optics were redesigned.  The first implementation of the LO optics used two separate LO chains, having output signals whose polarization is perpendicular to each other and combined through a polarizer grid, in order to propagate both LO signals on a nominal common axis.  Then, on the main optics plate, another polarizer grid would divide them again while being combined with the corresponding RF signals for each polarization.  When using one LO chain at a time, illuminating its corresponding subarray worked fine, but when operated in parallel, severe interferences between the two LO systems occurred. The solution was to decouple completely the two LO paths, keeping one on-axis (for the so-called H-polarization subarray) while the other LO had a tilted axis.  Therefore the whole LO and LFA optics plate was redesigned to accommodate this change, which proved to be successful as the strong interferences disappeared.  Fig. \ref{LFA_LO_diagram} shows the updated diagram of the LO chains. 
  
 \subsection{LO limitations and LO improvements}

The upGREAT/LFA receiver is designed to cover the 1.8-2.5 THz range. As such, the HEB mixers and the optics are all designed to cover that frequency range.  Currently, the upGREAT/LFA system is offered to the community in the 1.83-2.006 THz range.  This frequency limitation comes from the local oscillator technology. The VDI local oscillator multiplier sources allow to cover a moderate region around 1.9 THz with very high output power.  This development proved to be very challenging, and only the range 1.83-2.006 THz could be covered with several  local oscillator AMC chains, and only one VDI system was capable to extend as far as 2.07 THz. This is especially important, as there are several astronomical lines of interest in that upper frequency range.  Several attempts to replicate this performance with other AMC chains have not been successful until now.  Therefore observations in the 2.006-2.07 THz range can only be offered through consortium flights, shared-risks, until more funding is allocated for more robust, additional LO sources. 

Another major change was the redesign and re-arrangement of the LO components inside the LO main enclosure.  Indeed, it was found out that the thermal dissipation and air flow was not sufficient to adequately cool the main LO components, especially the power amplifiers from Spacek Labs. Each LO system generates a lot of power, 50 W at 26 GHz, 250 mW at 220 GHz, and 20-30 {$\mu$}W at 1.9 THz (or 40-60 {$\mu$}W when the last two triplers are cooled down to 100 K). They dissipate each 50 W, then the Stirling cooler for the triplers also dissipates around 100 W, therefore all in all, about 300 W are dissipated inside this small LO enclosure. Under normal laboratory conditions, at 20$^{\circ}$C, the power amplifiers case temperature  would be around 35-40$^{\circ}$C. But once the instrument is installed onboard SOFIA, particularly during summer months in California's Mojave Desert, it has proven difficult to manage the cabin temperature in the hours leading up to take-off.  Cabin temperatures approaching 30$^{\circ}$C in the aft cabin area can result in component instability and even failure.  To improve this situation, components were rearranged, the air flow was improved, power amplifiers are now cooled conductively using heat sinks with cooling pipes together with more powerful fans.   These changes resulted in $\sim$10$^{\circ}$C lower temperatures in the LO enclosure. Though these design changes certainly improved the thermal overtemperature situation within the LO enclosure, the stability of these components is still susceptible to large cabin air temperature fluctuations in the first few hours of a flight.  

There were also relatively strong interferences, spurs, present in many places in the RF band, originating from the LO chains. VDI has been optimizing and improving on the situation since 2015.  Several iterations of multiplier doublers have helped to clean up most of the band.  A major breakthrough in this regard occurred at the of 2017, when VDI identified a source of interference causing a forest of narrow spurs evenly spaced around 1 GHz IF (at the IF output of the receiver). These narrow spurs had been present from the beginning for all pixels, and were flagged during data reduction.  They originated  from pickup from the multipliers bias lines of a small part of the input RF signal (via radiation). A redesign of this VDI electronic controller is ongoing, but the short term solution was to include 10 MHz low pass filters for every multiplier bias line. This effectively solved the LO-based issues (while observations may still suffer from SOFIA-internal but also externally rooted RFI). 

The most recent improvement performed in 2018 was the addition of LO attenuation wire grids, placed at the LO beam outputs (see Fig. \ref{LO1}), in order to independently adjust the LO levels for both sub-arrays.  Before that, LO adjustment was performed electronically, or not done at all, therefore occasionally resulting in overpumped mixers, with some degradation in performance.

\section{HFA array}

The second upGREAT array for high frequencies (HFA) was completed in mid-2016, and its commissioning started in November 2016.   

\subsection{ Initial design, optics }

As described in \citet{risacher2016a}, the cryostat was fabricated at CryoVac GmbH, integrated with a transMIT pulse tube 2-stage cooler (PTD-406C).  The cryostat design is identical to the lower frequency array system (LFA), with some differences: 
\begin{itemlist}
\item Cryostat Window :		1 mm thick Silicon with Parylene-C coating and 525 ${\mu}m$ Silicon with anti-reflection grooves \cite{wagner2006}
\item Infrared Filters:		Originally a low pass 8 THz filter from QMC was used at the 77 K stage, but it was removed as cooling power is sufficient.  When removed, the impact on the cryostat and mixer temperatures was negligible. The loss at 4.7 THz of that filter was of ~7 percent.
\item Coupling optics: The optics of the upGREAT/HFA are based on the optics design for the LFA.  An assembly of six offset-parabolic mirrors in front of the six offset pixel HEBs forms the focal plane unit inside the cryostat. A dedicated elliptical mirror for the mixer of the center pixel is mounted separately behind this unit. Four active mirrors – common to all pixels and forming two Gaussian telescopes - map the cryostat focal plane to the focal plane of the SOFIA telescope. The cryostat houses one of these active mirrors, the other three are placed on the warm optics plate inside the optics compartment of the instrument structure. A first set of optical components was manufactured on an in-house Kern ultra-precision milling machine and are estimated to have a surface accuracy of a few microns RMS. Switching from these bare aluminum mirrors to a commercially manufactured set of gold-plated mirrors with a specified surface accuracy of 10 nm did not noticeably improve the instrument performance.
On the warm optics plate, the local oscillator beam and the signal beam are superimposed by a thin Mylar foil (2.5 to 3.5 ${\mu}m$ thickness). A 3.3 THz low-pass dichroic foil manufactured by QMC is used to split the HFA signal beam and the beam for the second installed GREAT channel. The optics are prepared for implementation of a second 7-pixel subarray for dual polarization observations. 

\end{itemlist}

%
%
%
%

\subsection{ Mixers fabrication }

\begin{figure}[h]
\begin{center}
\includegraphics[angle=0,width=5in]{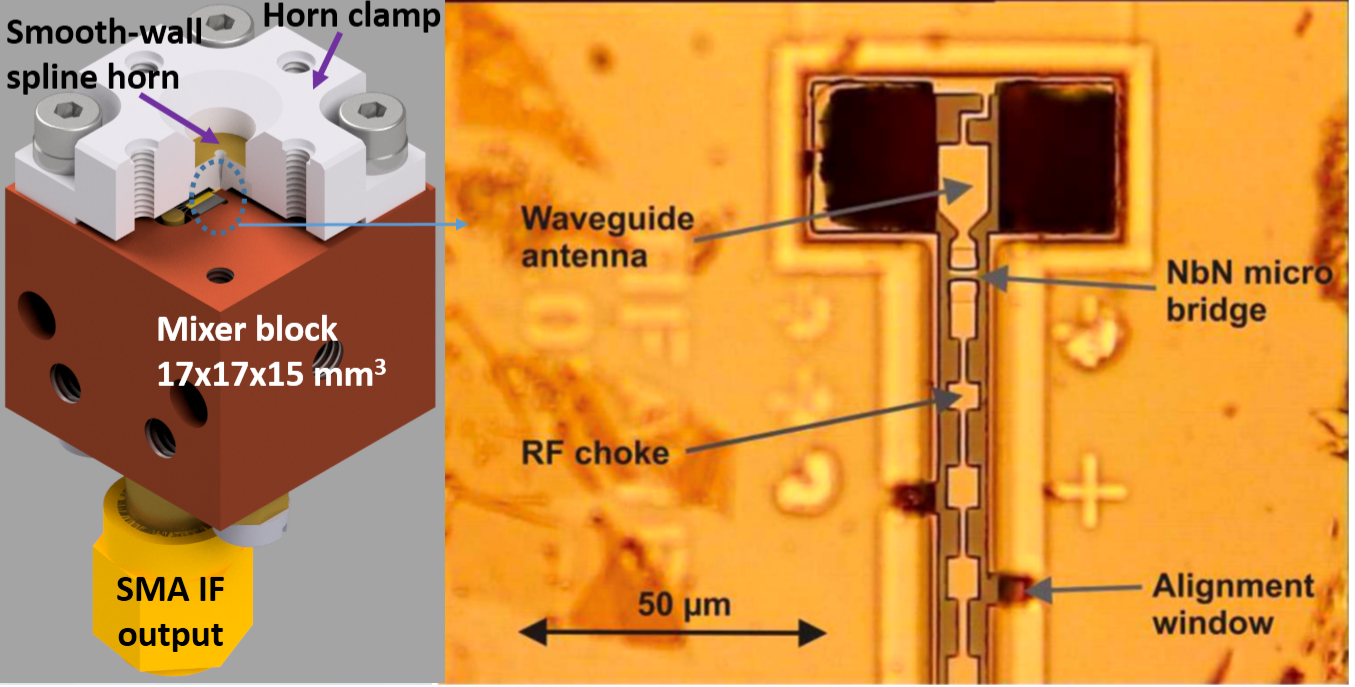} 
\end{center}
\caption{Left: cross section view of one of the HFA mixer blocks showing the main components. Right: View of the mixer chip, showing the input waveguide, the waveguide antenna and the NbN HEB.}
\label{hfa_block2}
\end{figure}

The HEB device fabrication, the feedhorns manufacturing and the subsequent integration of the mixer block proved to be challenging tasks.  The feedhorns used are based on the smooth-wall spline horns \cite{granet2004} and are a scaled version of the LFA horns.  They were fabricated by electro-forming by Radiometer Physics,  and they proved to be very difficult to fabricate with the required accuracy (low yield), mainly due to the difficult removal of the mandrel material inside the small waveguide section. Also in the block manufacturing in the Cologne workshop  and especially in the assembly tolerances the approximate factor of 2 reduction in all dimensions compared to the LFA made a noticeable difference. More devices became damaged upon assembly  and all mixers had to be pre-tested as single pixel in Cologne to ascertain their performance as a check on the assembly tolerances. The device fabrication was also not straightforward. As was done for the upGREAT/LFA mixer junctions, the HEB film thickness was decreased in order to reduce the LO power requirement for the array. However those reduced-thickness devices did not perform as well as the original 4.7 THz single pixel HEB  mixer that flew on SOFIA \cite{buechel2015}. Also, it appeared that the extremely thin NbN films suffered more from environmental effects resulting in performance deterioration over several cooling cycles. Finally, as there is sufficient local oscillator power with the QCL technology (see 5.3), the original HEB device was kept and the upGREAT/HFA devices are based on it, reproducing the previous single pixel performance. A cross section view of the HFA mixer block and chip assembly can be seen in Fig. \ref{hfa_block2}.

\subsection{QCL Local oscillators}

The local oscillator technology uses Quantum-Cascade Lasers (QCL). Two groups, the Institute of Optical Sensor Systems, German Aerospace Center (DLR) and the Physikalisches Institut der Universit\"at zu K\"oln, (KOSMA hereafter) have carried out parallel developments, and both have succeeded in providing flightworthy hardware. The DLR QCL has been flying since 2016 with the HFA array, and the KOSMA QCL started flying since May 2018. 

\subsubsection{DLR QCL LO evolution and performance}

\begin{figure}[h]
\begin{center}
\includegraphics[angle=0,width=4in]{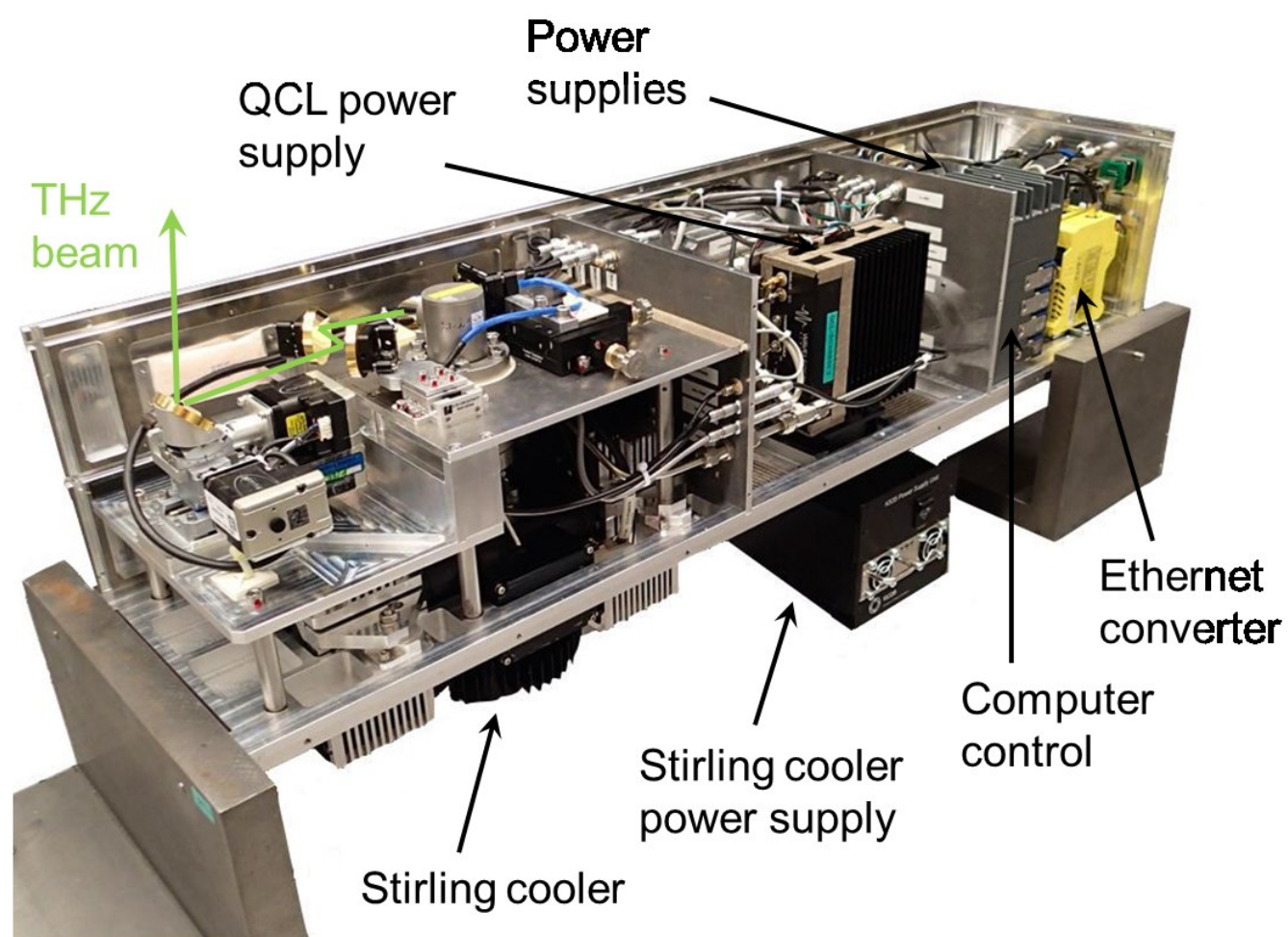} 
\end{center}
\caption{Photograph of the opened LO box on a holder in the laboratory. Dimensions of the box are 100 x 23 x 23 cm$^3$. The power supply for the Stirling cooler is the only external component which is needed for operation of the system}
\label{qcl0}
\end{figure}

    The DLR QCL system (see Fig. \ref{qcl0}) is a further development of the GREAT single pixel 4.7 THz receiver, which had been successfully operated on board of SOFIA from May 2014  to November 2016 \cite{richter2015}. The QCL LO system consists of a 4.7-THz QCL with a compact, low-input-power Stirling cooler (Ricor K535). The QCL is mounted inside a vacuum housing on the cold finger of the cryocooler, which provides a temperature between 30 and 80 K, necessary for the operation of the QCL. No liquid cryogens are required. The active medium of the QCL is based on a hybrid design and has been developed for continuous-wave operation, high output power, and low electrical pump powers \cite{schrokkte2013}. Efficient carrier injection is achieved by resonant longitudinal optical phonon scattering. This design allows for an operating voltage below 6 V. The amount of generated heat complies with the cooling capacity of the Stirling cooler of 7 W at 65 K with 240 W of electrical input power \cite{richter2010}. The QCL has a lateral distributed feedback grating, which is optimized for the transition frequency of OI at 4.745 THz. Due to the DFB grating the QCL has a single mode emission. The radiation is coupled out through one of the end facets of the single-plasmon waveguide and focused with a parabolic mirror. The beam profile is almost Gaussian with an M$^2$ value of approximately 1.2. A flat mirror which is mounted on a goniometer allows aligning of the QCL-beam in order to direct it to a Fourier grating to multiplex the single beam into 7 equal beams (see 5.4).
The emission frequency of the LO is determined by measuring the absorption spectrum of CH{$_3$}OH and comparing this with data from the literature. The LO covers a frequency range from -1.5 to +6.5 GHz around the rest frequency of [OI]. The frequency is tunable by changing the laser driving current as well as by changing the heat sink temperature. At a constant heat sink temperature the frequency coverage of the LO is up to 4 GHz, for example at 52 K it ranges from -1.1  to +2.6 GHz. In general, tuning of the frequency by current is preferred, because it is faster than tuning by changing the heat sink temperature, which requires more time, because thermal equilibrium has to be established. The heat sink temperature is kept constant within $\pm$ 1 mK by a dedicated control loop. The frequency stability is about a few MHz \cite{richter2015}. The output power varies with the laser driving current and with the heat sink temperature. For all LO frequencies it is above 1.2 mW with a maximum power reaching 2.2 mW. This includes absorption loss by the window in the vacuum housing of the Stirling cooler as well as atmospheric absorption loss. Corrected for these losses the output power of the QCL is up to 6 mW. The LO system is fully controlled by a dedicated software which controls the Stirling cooler itself, the electrical input power of the QCL, and the heatsink temperature.

\subsubsection{KOSMA QCL LO performance}

\begin{figure}[h]
\begin{center}
\includegraphics[angle=0,width=4in]{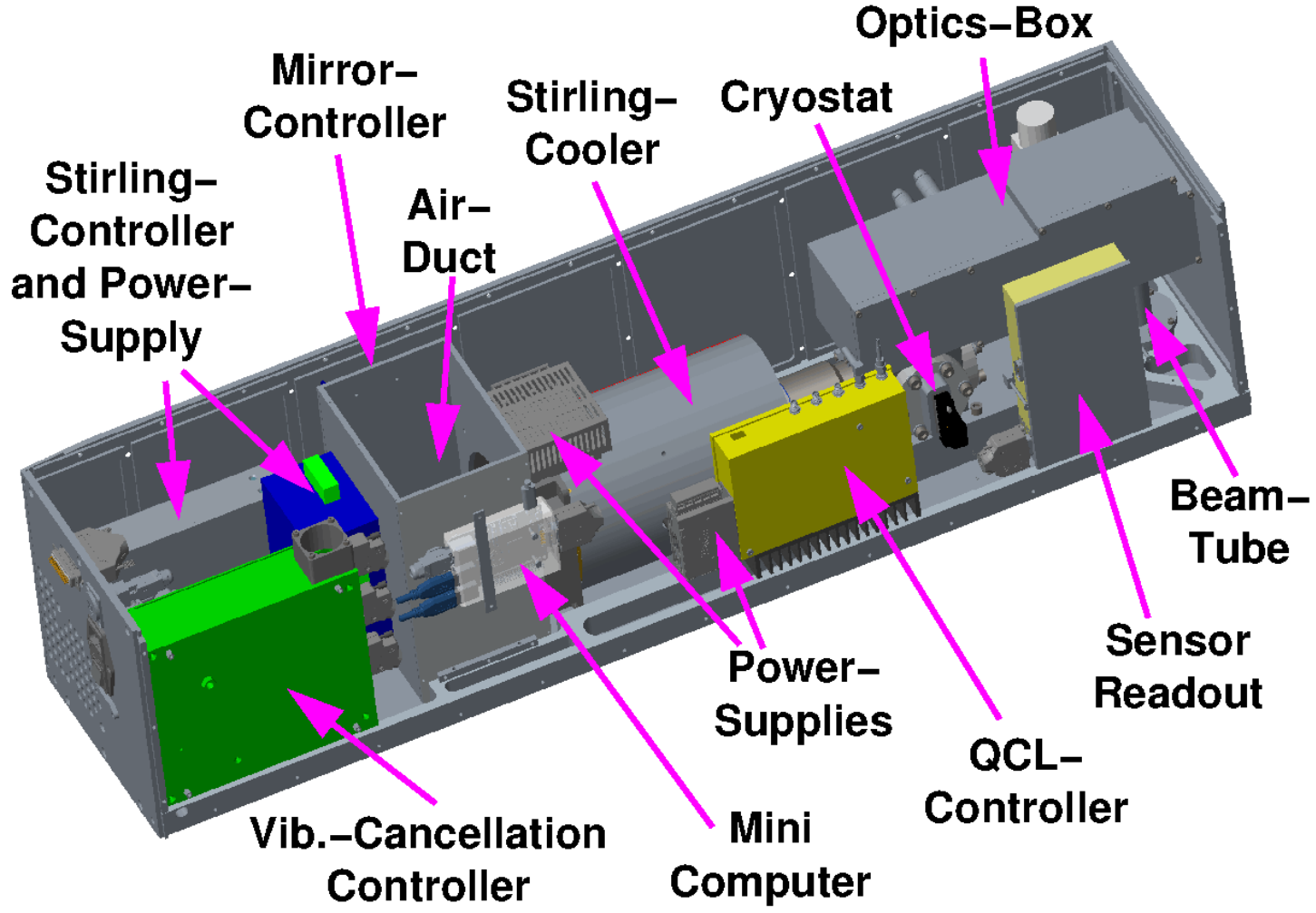} 
\end{center}
\caption{KOSMA QCL LO box enclosure showing the main components. It has the same outer dimensions as the box shown in Fig. \ref{qcl0}.}
\label{qcl1}
\end{figure}

\begin{figure}[h]
\begin{center}
\includegraphics[angle=0,width=4in]{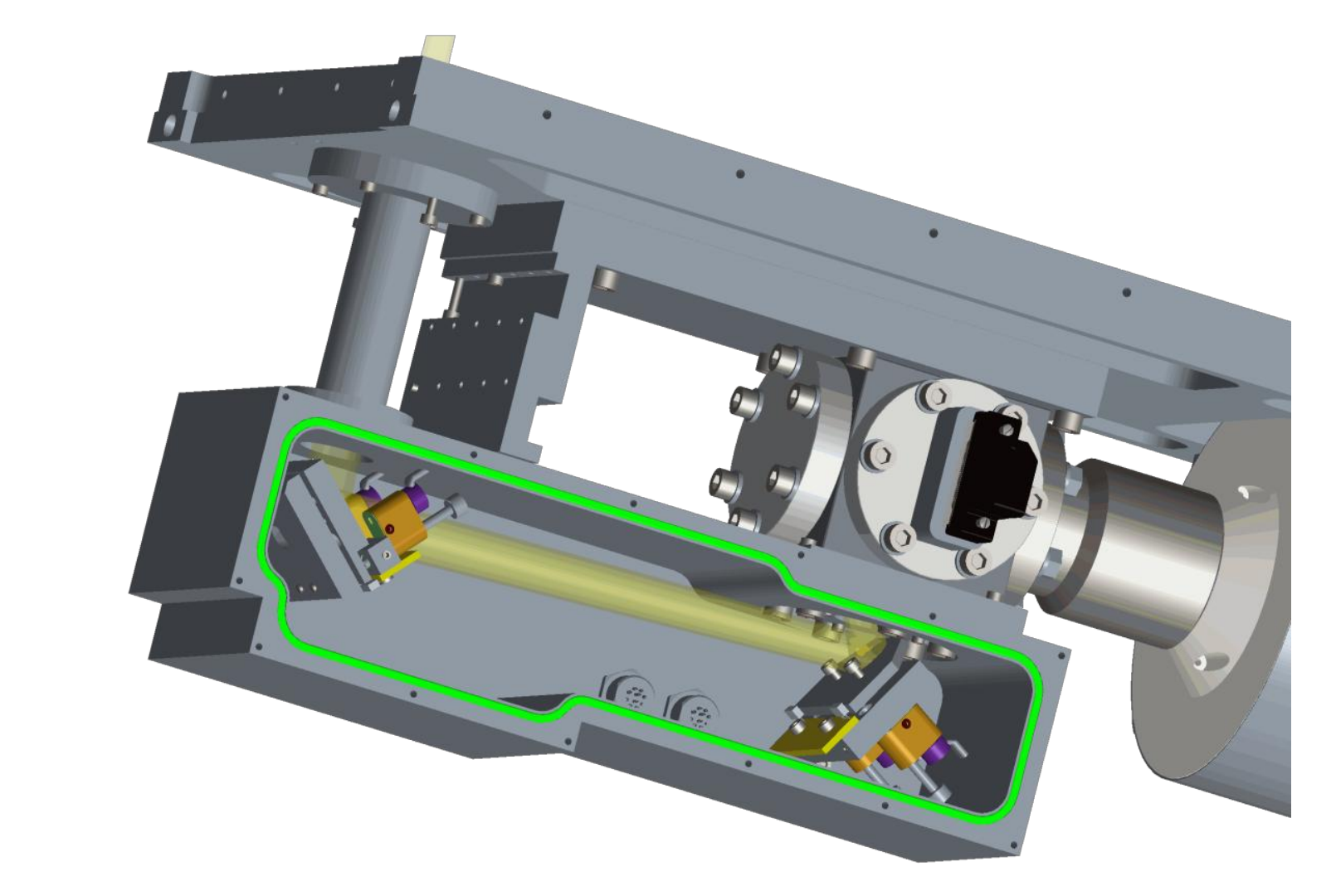} 
\end{center}
\caption{The KOSMA QCL optics is located inside a hermetic box indicated by the green contour. This box is connected to the main instrument compartment via sealed fittings, without any RF windows (as opposed to all the other GREAT local oscillator boxes). This allows during flights, to have the QCL optics compartment at the ambient outside air pressure, about 160 mBar, therefore maximizing the QCL LO signal transmission}
\label{qcl2}
\end{figure}

The KOSMA QCL LO (see Fig. \ref{qcl1} and Fig. \ref{qcl2}) relies on a single mode, double metal QCL with integrated patch antenna array \cite{bosco2016, justen2016} delivering about 2.4 mW optical power in the fundamental Gaussian mode. About 25 to 30 percent of the power is needed to pump the 7 pixel HEB mixer array. The power can be varied by a polarizer grid in the HFA optics compartment. The QCL can be tuned -2.5 to +2.9 GHz around the OI line at 4744.8 GHz by controlling the bath temperature to a few mK between 40 and 70 K. The QCL is cooled by a Stirling cryocooler. No liquid cryogens are required. The beam position and beam angle are adjusted with two tip-tilt mirrors which couple the beam into the HFA optics compartment. In order to avoid absorption of the LO signal by water in the surrounding air, the beam path is contained in a hermetic housing, which is open to the inside of the instrument optics compartment and therefore experience during flight the same external air pressure (typically about 160 mBar). The LO system is controlled remotely over an Ethernet connection by a dedicated computer which controls the Stirling cooler, vibration control, QCL current source, QCL temperature control and mirror positions.  A prototype of this QCL was used in Cologne to pre-test the 4.7 THz mixers in a single pixel set-up. 

\subsubsection{QCL LO frequency stability}

Currently, the QCL LOs are free-running and their absolute frequency stability is set by the temperature and current stability of the lasers. For the upGREAT/HFA observations, the [OI] frequency scale is established by referencing to the telluric [OI] that is prominently present in all data, to a precision of better than 0.1 km/s, however this requires additional post processing steps in the data reduction. For the DLR QCL, the estimated linewidth is of a few MHz and drifts can be of the same order. When using the KOSMA QCL, measurements of the beat signal between the QCL and a microwave source showed that the QCL frequency jitter is smaller than 1 MHz (0.05 km/sec). Similarly, the laser frequency drift during observations, as measured by comparison with the telluric OI absorption line is also better than 1 MHz. Several options are under study to achieve a frequency/phase stabilization of the QCL, the main obstacle for this being the limited physical space available in the instrument to place additional components.

\subsection{ LO distribution}

\begin{figure}[h]
\begin{center}
\includegraphics[angle=0,height=2.2in]{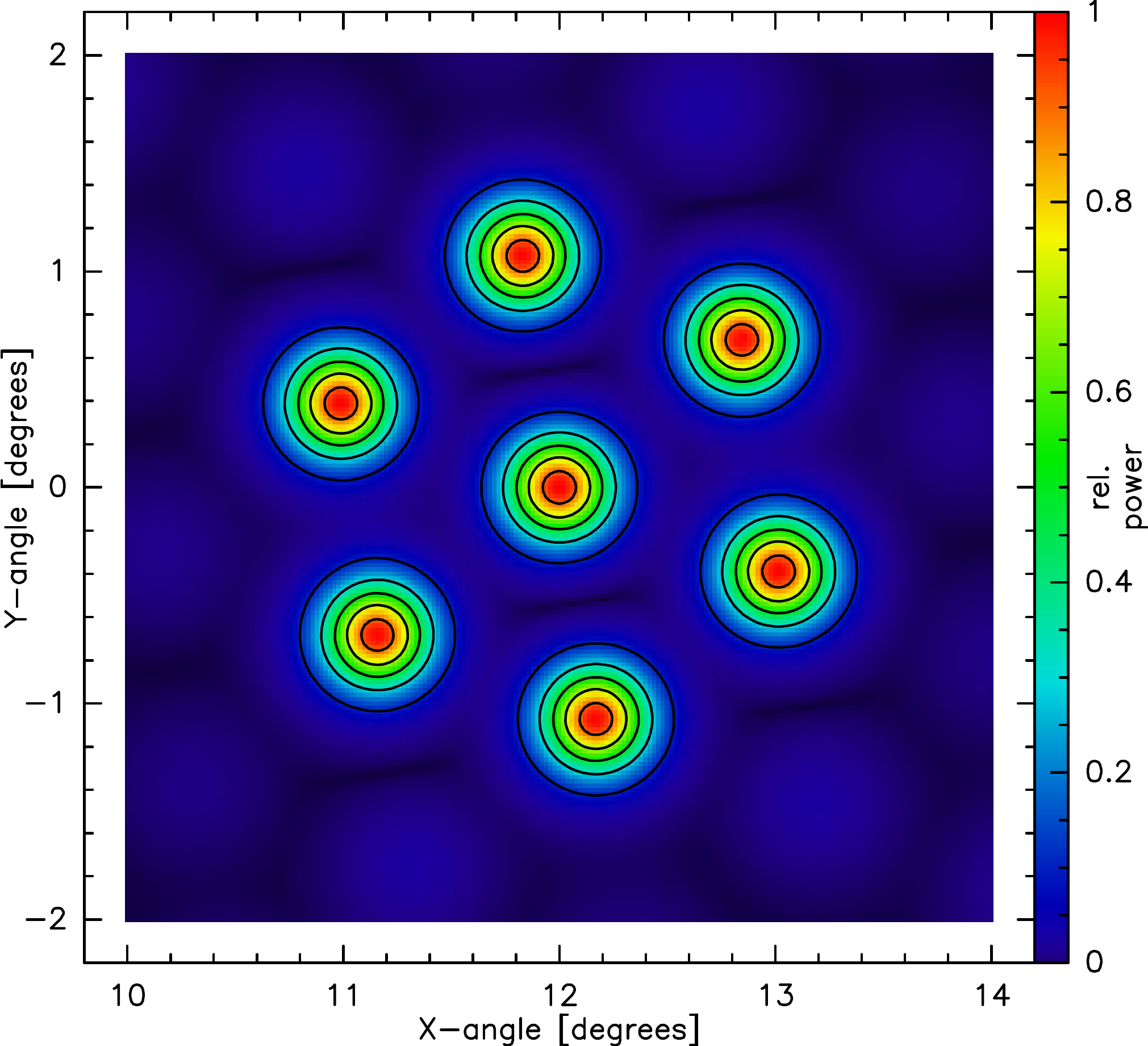} 
\hfill
\includegraphics[angle=0,height=2.2in]{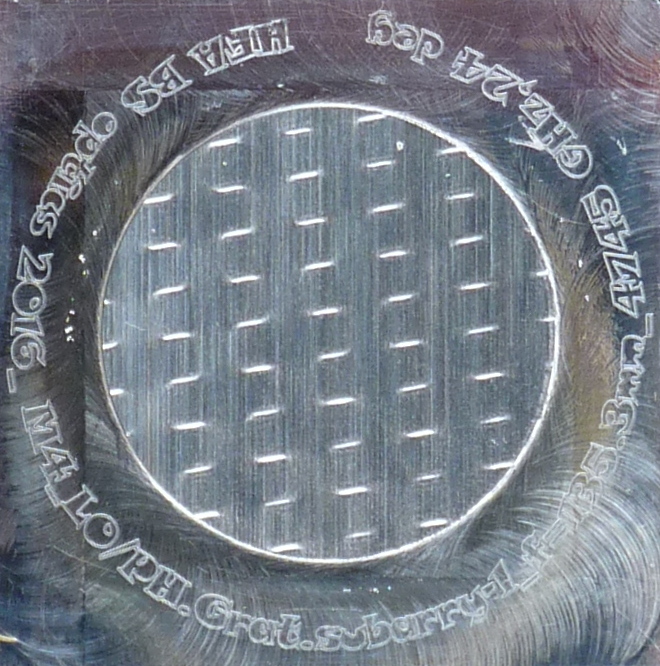} 
\hfill
\includegraphics[angle=0,height=2.2in]{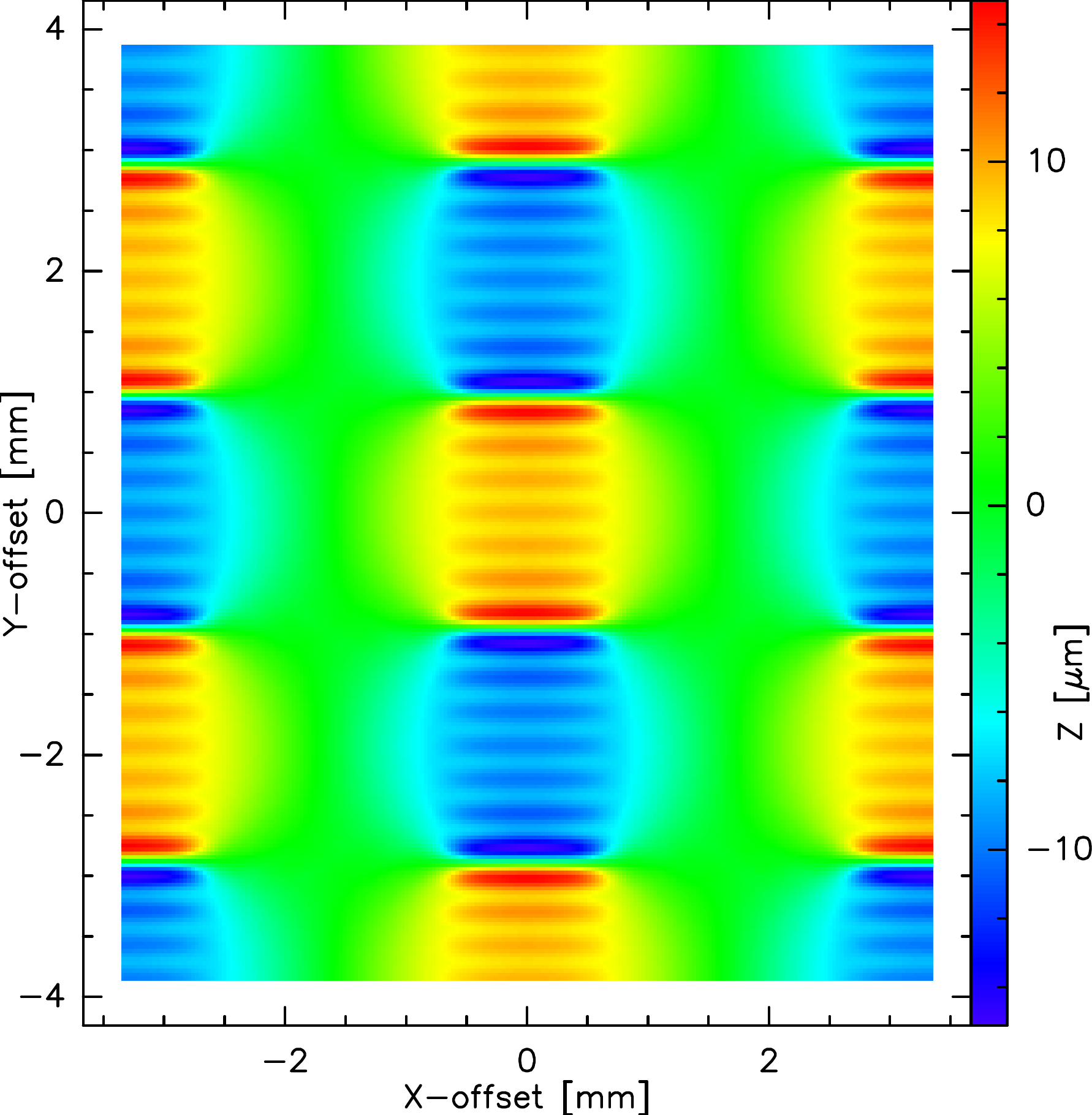} 
\end{center}
\caption{Collimating Fourier grating operating at 4.7 THz: calculated 
diffraction pattern (left), photograph of the final grating (center) and 
surface structure (right). We show two unit cells of the grating surface to
illustrate its hexagonal symmetry.}
\label{CFG}
\end{figure}

For the LO distribution, as in the LFA array, one single LO source is equally multiplexed into seven beams via a phase grating. This collimating Fourier grating (CFG), distributes the LO signal to the array elements \cite{heyminck2001}. The grating structure (shown in Fig.  \ref{CFG}) is the same as in the LFA band \cite{risacher2016a} or in the CHAMP+ receiver \cite{kasemann2008}. The nominal efficiency\footnote{calculated as the fraction of power contained in the desired diffraction orders relative to the total power in the diffraction pattern} of the grating is approximately 90\%. The RMS power imbalance between the seven beams is ideally less than 0.5\% at the designed frequency. The CFG has a diameter of 20 mm, a focal length of 135.3 mm and operates at a reflection angle of 24$^\circ$. In order to comply with opto-mechanical boundary conditions, the grating structure is designed to produce a diffraction pattern rotated by 9$^\circ$ with respect to the reflection plane.

\section{Instrument installation onboard SOFIA}

\begin{figure}[h]
\begin{center}
\includegraphics[angle=0,width=4in]{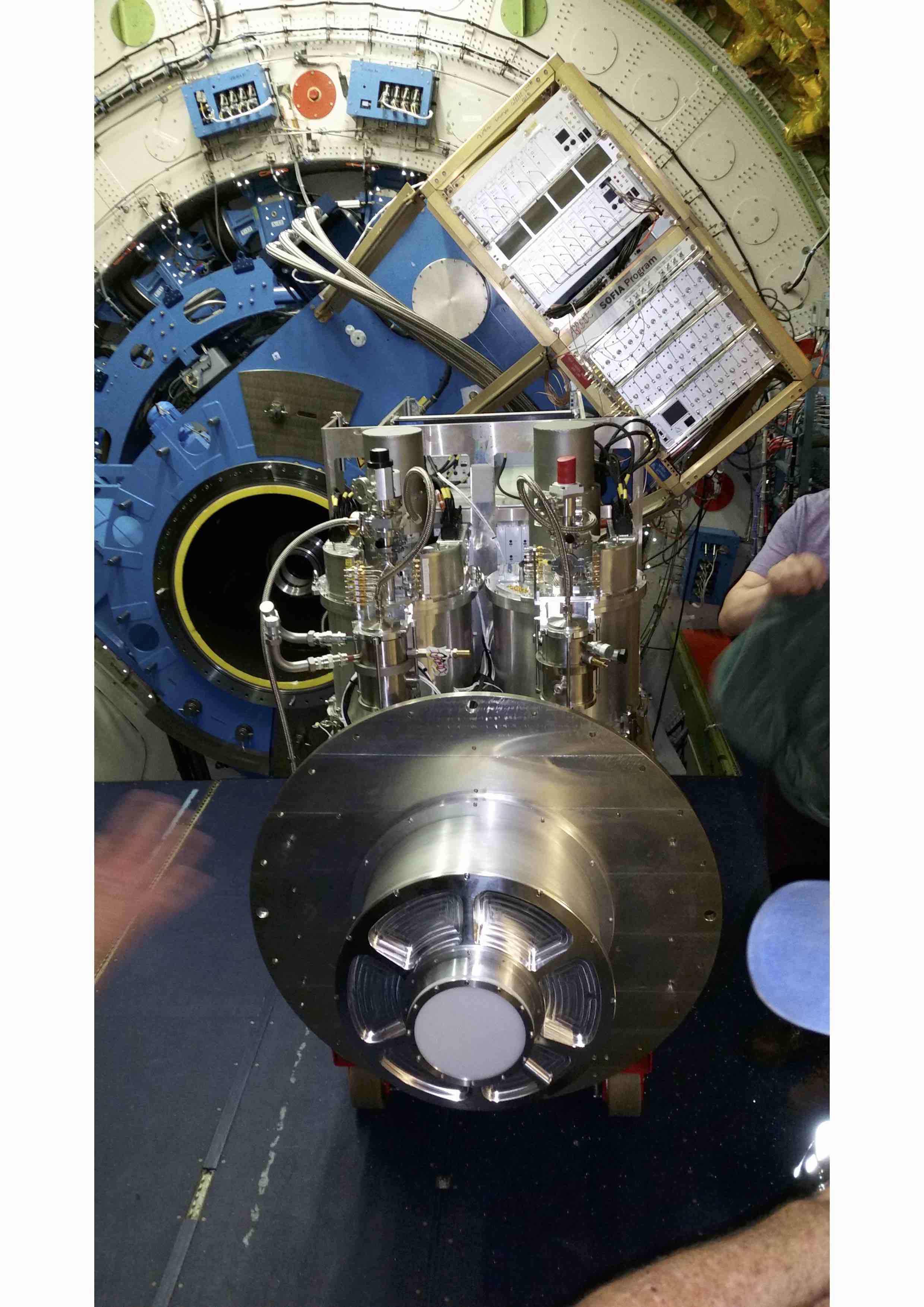} 
\end{center}
\caption{Science Instrument before being turned by 180 degrees and connected to the interface flange (with the yellow ring). The HDPE window visible at the front of the instrument is removed before the final connection (only used for laboratory characterization to allow pumping the optics compartment)}
\label{SI_install}
\end{figure}

This section describes the setup and typical installation procedure of the Science Instrument (SI) into the aircraft. The various SOFIA science instruments are connected to a main reference, the SI flange, which is at the end of the Nasmyth Tube.  This is inherited from the Kuiper Airborne Observatory (KAO), SOFIA's predecessor, with a 36 inch diameter primary mirror on a Lockheed C-141A platform which was operational from 1975-1995.  Typically, an instrument is installed aboard SOFIA and stays there for a flight series lasting from one to several weeks.  Figure \ref{SI_install} shows the GREAT instrument with the upGREAT arrays on the aircraft just before the final installation and connection to the SI flange.

\begin{figure}[h]
\begin{center}
\includegraphics[angle=0,width=5in]{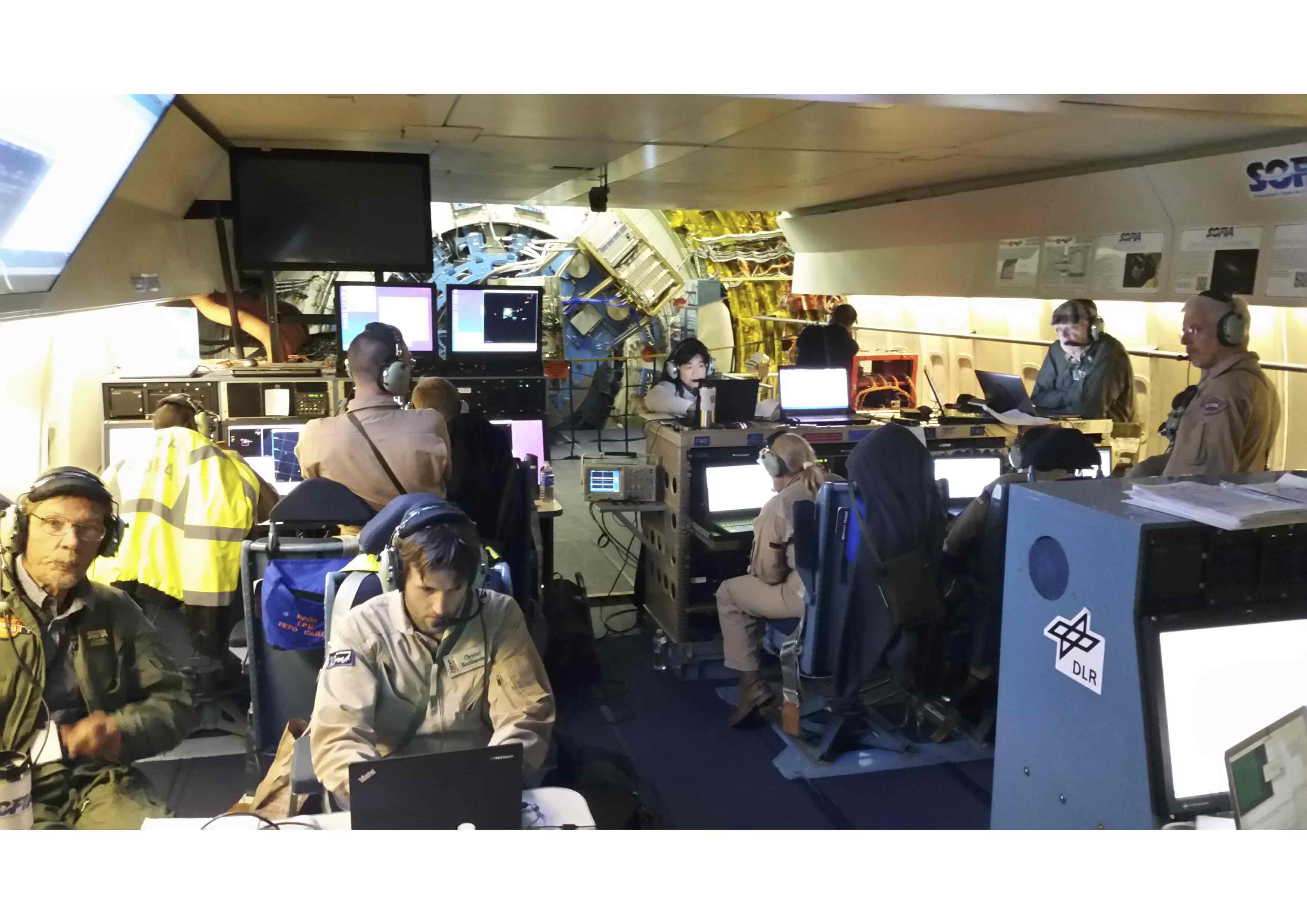} 
\end{center}
\caption{ Observers onboard SOFIA during a flight using the GREAT instrument. The science instrument is located at the far end (the counterweight rack with its instrumentation can be seen at the aft end of the pressurized main cabin). The instrument team can be seen on the right side of the photograph, tuning the instrument, sending the observing commands to the telescope and analysing the collected data in real-time. On the left of the picture, the telescope team ensures the pointing of the telescope using its various star trackers, and monitors the good operation of the telescope during the observations.  On the far right side, a small portion of the mission director (MD) console can be seen.  The MD is the responsible for  coordinating all the teams together, the instrument team, the telescope team and the pilots.  }
\label{SI_obs}
\end{figure}

After the science instrument is connected mechanically, all of the electrical connections are done and tested. Typically, as the instrument has been undergoing testing in the laboratory facilities before the installation, it is already cold. The helium lines are connected from the instrument to the patch panel, and the compressors are started as soon as possible to resume its cooling.  As long as the SI installation can be done in under 2 hours, this procedure allows adequate time for thermal recovery of the closed-cycle systems, resulting in an operational system within half a day. Figure \ref{SI_obs} shows the typical work configuration during ground testing or flights.

\subsection{Cryocoolers  onboard SOFIA}

The cryocooler system is permanently installed as part of the observatory on SOFIA and includes components located primarily on the aft upper deck but with a remote control panel located in the main cabin. The cryocooler system provides closed-loop cryogenic cooling for science instruments with the following benefits:

\begin{itemlist}
\item Continuous cooling of the instrument as opposed to using expendable liquid cryogens, which typically have $\sim$24h hold time
\item Eliminates material handling hazards and cost of expendable cryogens
\item Simpler maintenance, operations, and logistics by eliminating the need to refill cryogen reservoirs
\end{itemlist}

\begin{figure}[h]
\begin{center}
\includegraphics[angle=0,width=6in]{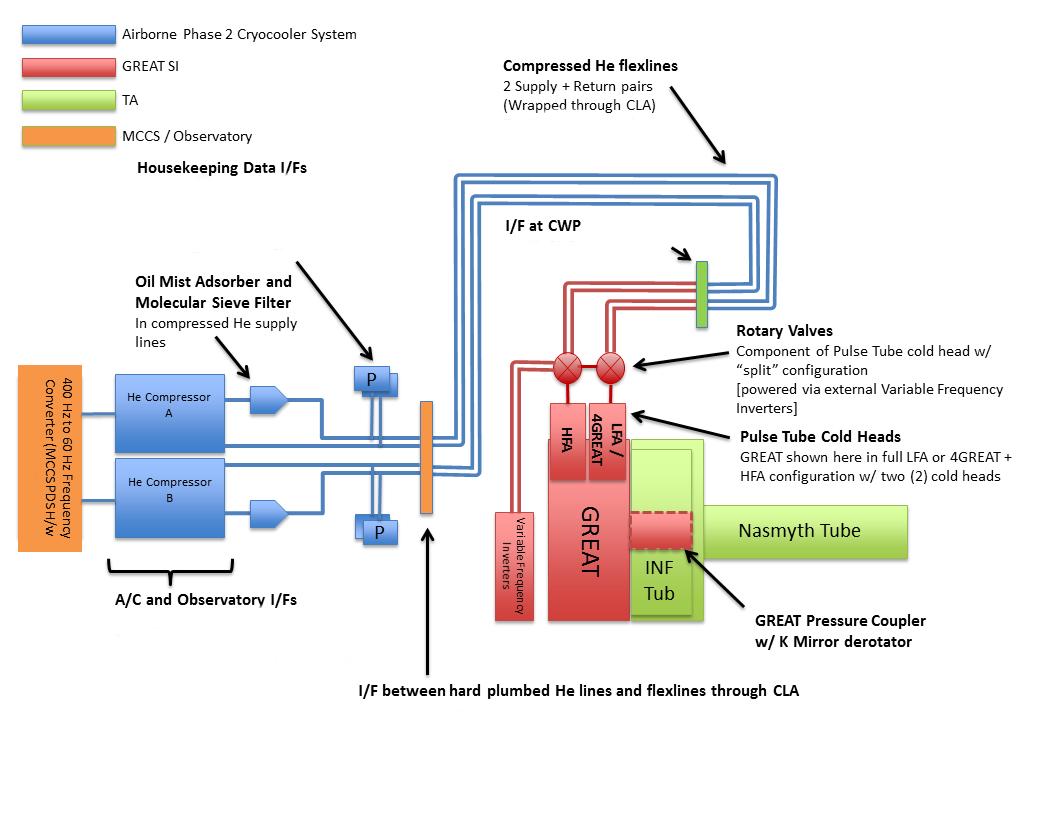} 
\end{center}
\caption{Schematic showing the SOFIA cryocooler infrastructure}
\label{cryocooler1}
\end{figure}

\begin{figure}[h]
\begin{center}
\includegraphics[angle=0,width=6in]{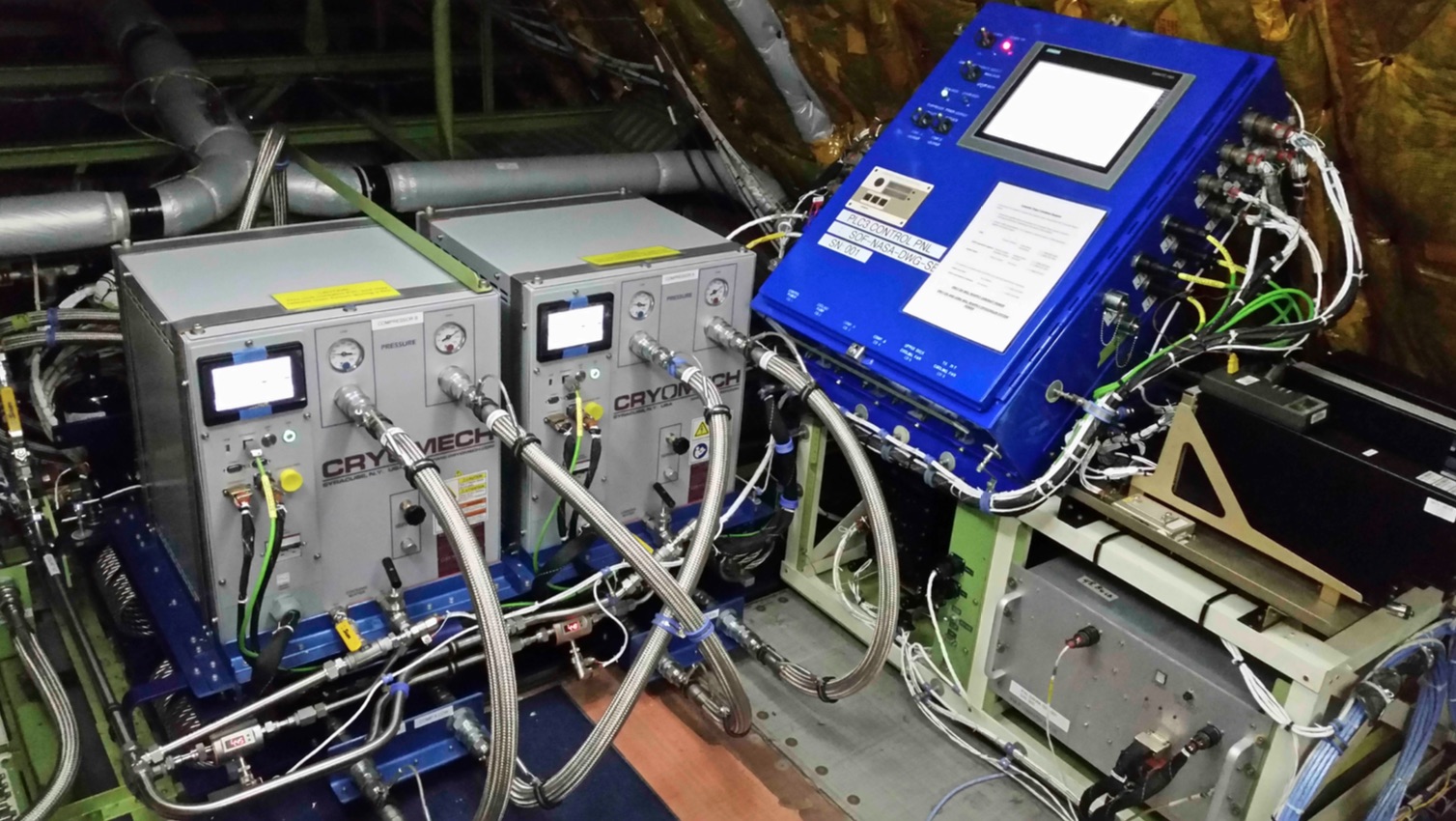} 
\end{center}
\caption{Compressors located on the upper deck of the SOFIA Boeing 747.}
\label{cryocooler2}
\end{figure}

The cryocooler system capable of driving two cold heads, completed Integration and Test (I{$\&$}T) and Verification and Validation (V{$\&$}V) in 2017 (overview shown in Fig. \ref{cryocooler1}, photograph in Fig. \ref{cryocooler2}). It consists of two ruggedized Commercial Off The Shelf (COTS) liquid-cooled scroll Cryomech compressors that use helium as an operating fluid to cryogenically cool the science instrument. Operation of the cryocooler is largely automatic, with two modes, a ground mode and flight mode that determine where and how the waste heat from the compressor(s) is rejected. The cryocooler has automatic monitoring and is designed to shut itself off in the event of critical faults.
The cryocooler draws about 23 kVA from the observatory bus 4, 115 VAC, 400 Hz, 3-phase power source, when both compressors are operating at maximum load. Eight frequency converters in the aft upper deck convert the input power from 400 Hz to 60 Hz for the compressors, coolant pump and controller operation. The two heat exchanger axial fans operate from unconverted 400 Hz, 3-phase power from observatory bus 4.

The cryocooler power source is downstream of the cabin depressurization relay which will remove power from all components in the event of depressurization.

To maintain ideal science instrument detector temperature, the cryogenic cooling system should operate continuously, although short (30-45 minute) interruptions are tolerable and occur regularly when the plane is towed between the hangar and the airfield, or during maintenance. 

\section{Performance on sky }

The performance of the upGREAT/LFA system was already shown in \citet{risacher2016b}. We present here some results for the upGREAT/HFA system, taken during flight campaigns in 2017 and 2018.

\subsection {In-flight instrument performance verification}

\subsubsection {upGREAT/HFA receiver sensitivities and stability}

\begin{figure}[h]
\begin{center}
\includegraphics[angle=0,height=2.6in]{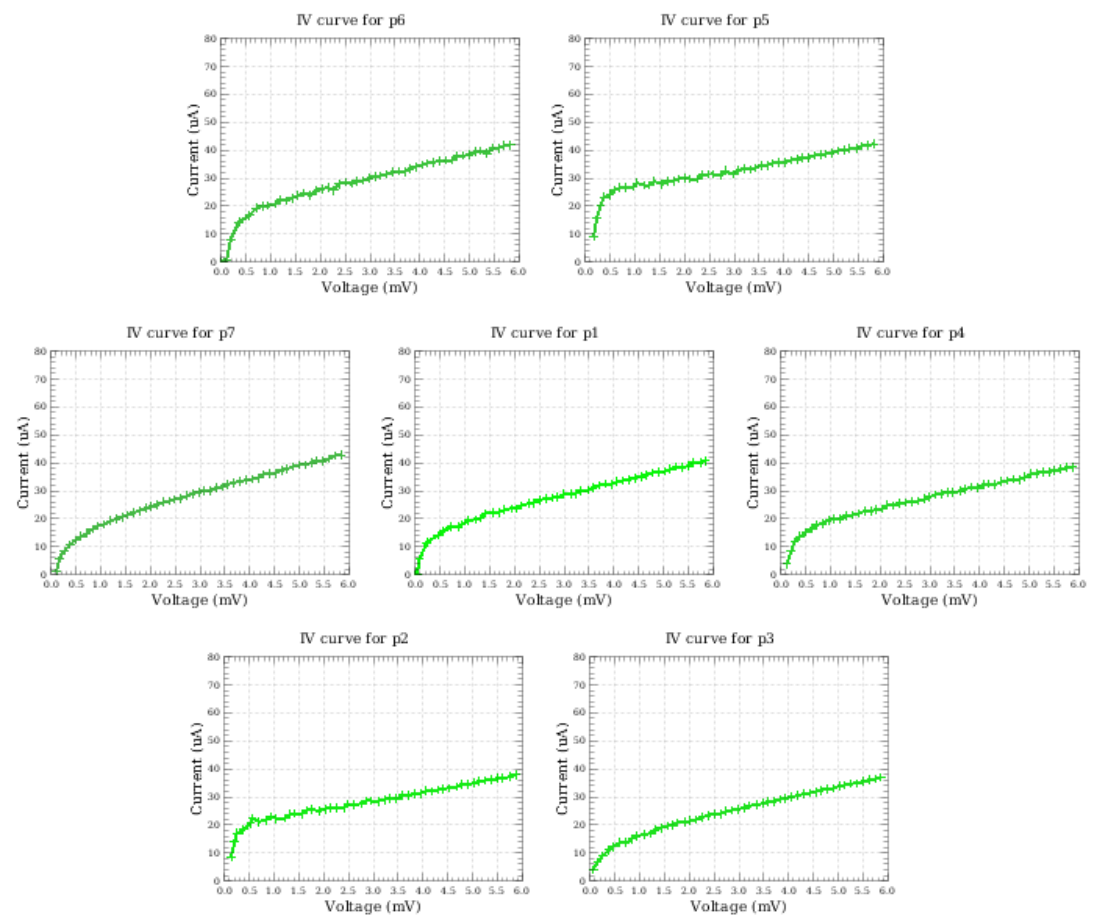} 
\hfill
\includegraphics[angle=0,height=2.4in]{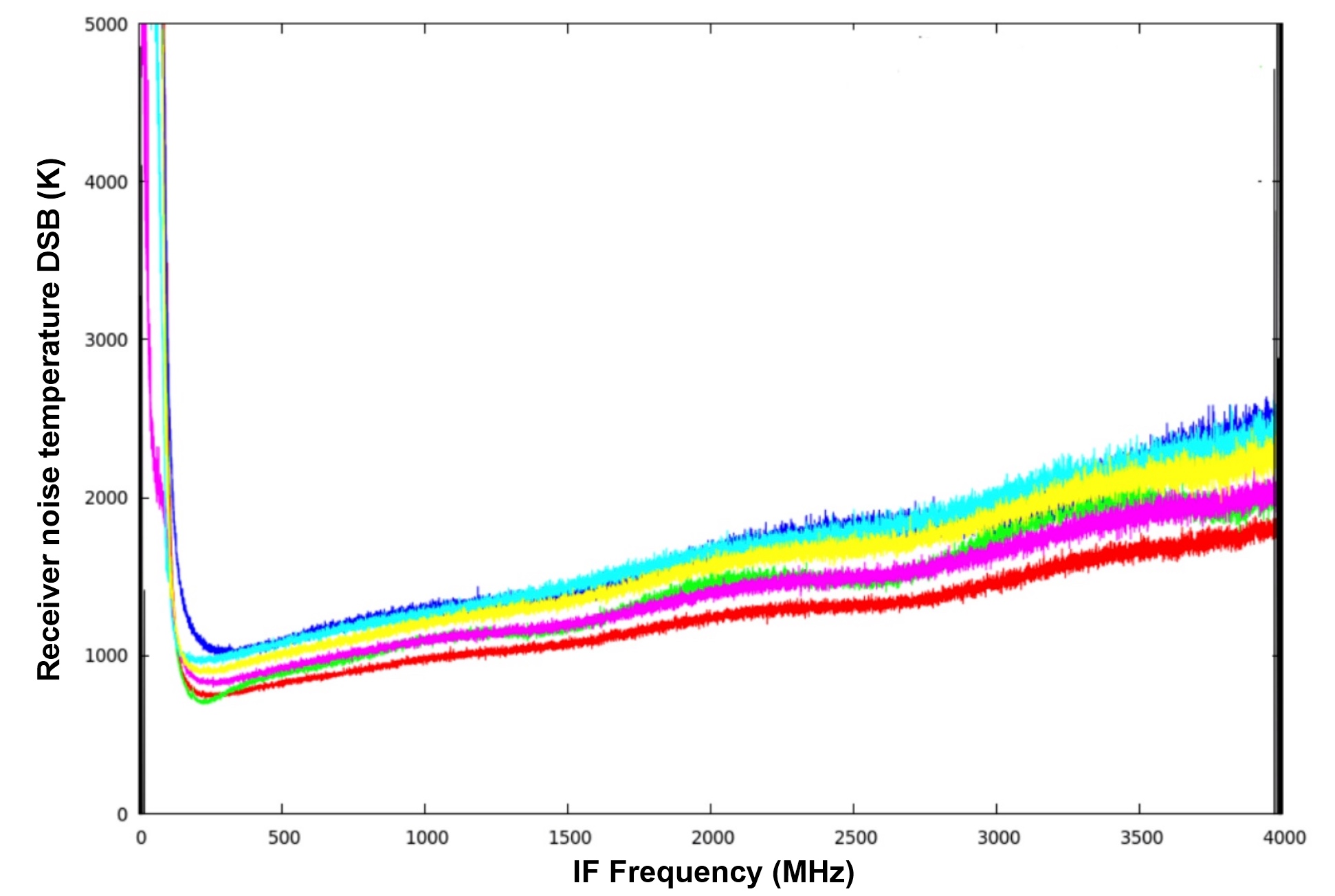} 
\hfill
\end{center}
\caption{Left: IV curves with LO power showing evenly balanced LO power for all pixels. Right: Double sideband noise temperature for the 7 pixels when pumped under the best conditions (black and red curves are on top of each other). This represents the best achievable performance. Under normal flight conditions, as there is little time for tuning optimizations, the performance degrades for some of the pixels. }
\label{HFA_Trec}
\end{figure}

The upGREAT/HFA best achievable sensitivity can be seen in Fig. \ref{HFA_Trec}.  The left plot shows the HEB mixers IV curves for the best LO distribution between pixels. The mixers achieve their optimum performance only over a narrow range in LO power, which is not identical for all pixels. The right plot shows the measured double sideband  receiver noise temperature over its IF range with bias settings of 1.5 mV for all pixels. For projects requiring best stability, different bias settings are recommended, with slightly degraded sensitivities. During normal observations, the time allocated for tuning optimizations can be less than a few minutes, therefore there is generally insufficient time to ideally set the parameters (the most critical one is the LO power distribution between the various pixels) and achieve the performance shown in Fig. \ref{HFA_Trec}. As a consequence, some of the pixels can have degraded performance. Thus, for each flight leg, the project scientist has to trade-off between observing time and best possible performance.

\begin{figure}[h]
\begin{center}
\includegraphics[angle=0,height=5in]{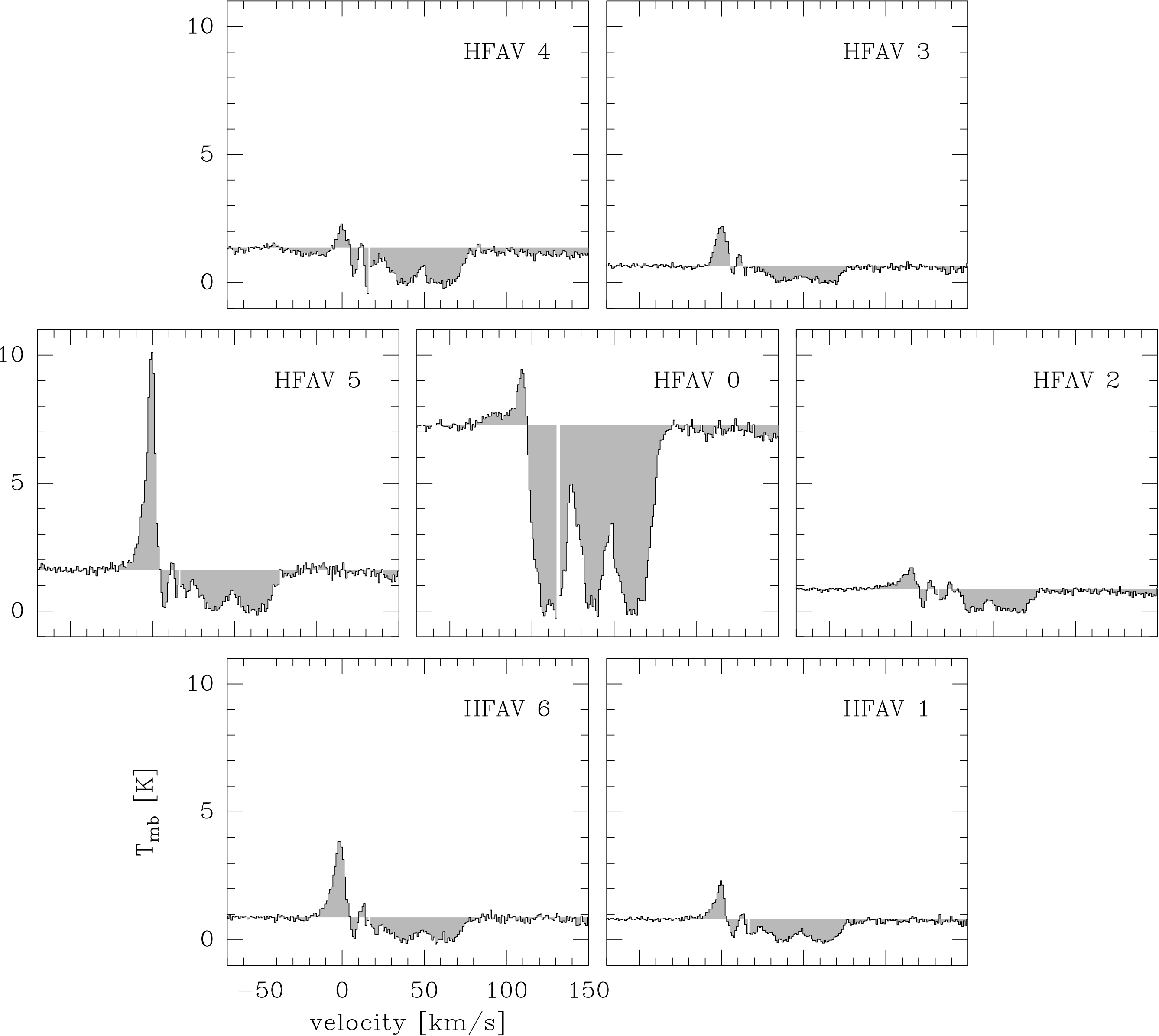} 
\end{center}
\caption{OI 63 $\mu$m spectra observed with the upGREAT/HFA towards the high-mass star forming core W49N, showing emission and strong absorption along the
line-of-sight towards the prominent dust continuum emission of the source. The data was acquired on SOFIA flight 420 out of Christchurch, New Zealand, on July 15 2017, during a 1.2 hour flight leg at an altitude of 43 kft. Observations were performed in double-beam chopped mode, wobbling the subreflector at 2.5 Hz with a throw of 180". No baseline was removed, the spectra were box-smoothed to a spectral resolution of 1 km/s. }
\label{HFA_spectra}
\end{figure}

To assess the instrument stability performance, a typical figure of merit is the Allan variance. The results for the upGREAT/LFA receiver were detailed in \citet{risacher2016a}, in section IV.B.4. 
 Under best stable ambient temperature conditions, the spectroscopic stability between 1.4 MHz width channels is typically above 80-100 seconds (Allan time). But if ambient temperature is not stable and drifts for example more than 1 degree per hour, the stability can easily degrade to 30-40 seconds. For the upGREAT/HFA receiver, performance is similar, only slightly worse, with spectroscopic Allan times typically better than 60 seconds.  It also appears that the upGREAT/HFA system is more sensitive to the pulse-tube induced mechanical vibrations, which modulate the LO coupling causing bandpass changes and ultimately the baseline can be dominated by the standing wave originating from the mismatch between the HEB mixers and the cryogenic amplifiers. There are several ways to mitigate this known effect. For chopped observations, data can be read-out synchronously with the wobbler phase, effectively suppressing this effect. With this solution in chopped mode, no degradation or limitation due to this effect is seen in the baseline quality up to integrations of $\sim$2 hours (more is difficult to accommodate with an airborne observatory).  An example is given in Fig. \ref{HFA_spectra} showing the typical achieved baseline quality. 
 
If total power observations are requested (e.g. on-the-fly modes, up to 20 seconds on phase), procedures have been developed to minimize this baseline modulation during post-processing (Higgins, in prep.). This takes advantage of the fact that the spectral pattern is constant for a given tuning (pixel-based) and hence can be fit to observations of “blank” sky (an approach similar to what has been done with Herschel/HIFI in \citet{kester2014}).

\subsubsection {Array parameters determination}

The necessary steps to be performed in the very first flight after (re-)installation of the instrument (or after a change of its configuration that might affect the optical alignment) are described for the LFA in \citet{risacher2016b}. For the upGREAT LFA+HFA characterization, we follow basically the same procedures, with parameters adjusted to the smaller (HFA) beam and the more compact array configuration: 

\begin{itemlist}
\item determine the instrument focal plane geometry (boresight and array geometry) 
\item determine the optimal focus position 
\item measure the beam coupling (efficiencies) on a known calibrator. 
\end{itemlist}

The first step is to determine the arrays geometry in the focal plane and hence their optical alignments to the optics of the telescope’s imager, which later on will control the tracking of the telescope (and hence the pointing of the GREAT instrument). Once the array geometry has been derived, the actual sky coordinate of each pixel will be introduced to the raw data header of each observation dump (typically every 0.3-1.0 sec).  
The alignment of, in particular, the HFA array (means, determining the pixel positions precisely) is challenging even on the brightest calibrators, requiring excellent residual atmospheric transmission. As an example, during our first and only flight out of Palmdale after re-installation for the OC6-G series on May 22$^{nd}$, 2018, flight planning enforced observations of Jupiter starting at low flight altitudes (36 kft) and mediocre atmospheric conditions (15 $\mu$m PWV). Short scale atmospheric instabilities (see Fig. \ref{atm2} for a display of the limited atmospheric transmission near the [OI] line) made it impossible to derive the HFA parameters, and only with our first southern deployment flight (cruising now at 41-43 kft in cold southern skies), the array parameters could be derived. 
From two observation of Mars in June 2018 (on June 16$^{th}$, the diameter of the planet was 18.9"), we determined the offset and angle of the rotator axis and the array geometries. We derive the individual pixel positions with an rms deviation of typically 1". The arrays’ central positions were co-aligned with respect to the central HFA pixel to 0.3“ (LFAV) and 1.0 ” (LFAH), so well tolerable within their 14.1” main beam.  In practice, for almost all projects we “tracked” on the central pixel of the HFA because of the higher tracking requirement of its smaller beam.

With extended Mars, the HFA beam sizes could not be well constrained during this year’s observations but as the internal array optics has not changed since the commissioning of the individual arrays in 2016 (when Mars was more compact, 7.6"), it is safe to adopt the beam sizes determined from those observations. 
On June 26$^{th}$, 2018, we derived the coupling efficiencies of the LFA and HFA arrays with dedicated cross-scans of the individual array pixels across Mars. In Table 2, we quote the average main beam efficiency per sub-array, with the standard deviation.  

\begin{wstable}[h]
\caption{upGREAT beam parameters (as determined on Mars in 11/2016 and 06/2018)}
\begin{tabular}{@{}ccc@{}} \toprule
   Parameter &  LFA (1.9 THz)  & HFA (4.74 THz) \\
 \botrule
Half power beam width   & 14.1"  &6.3"  \\
Array geometry: pixel spacing  & 31.8" & 13.6"  \\
Co-alignment of center pixel wrt. to HFA-V  &  0.3”(V), 1.0”(H) &    \\
beam efficiency (Martian disk: 18.9”) & 0.65 (H) / 0.66 (V) & 0.64 (V) \\
\botruleœ
\end{tabular}
\label{aba:tbl2}
\end{wstable}

There are several science applications for which the absolute pointing accuracy of the system (telescope plus instrument) is critical. In some cases of observations towards dark cloud complexes (a prominent example is IRAS16293) where there are no tracking stars available for the telescope’s focal plane imager and positioning based on gyroscopes and the wide field imagers may be unstable, drifting on short timescales to as much as 3-5", which is unsuitable for observations with a 6" beam.  Under nominal circumstances, with a bright star in the FPI field, optical tracking is good to 0.5 arcsec. Tracking the instrument on the central HFA pixel, additional uncertainty comes from the precision of the determinations of the rotator offset and axis and of the pixel’s offset from the rotator axis. The combination of these errors is obviously difficult to predict, but experience has shown that achieving an absolute positioning of HFAV00 to better than 1.5" is challenging. 

\subsection {Science use /operation }

\begin{figure}[h]
\begin{center}
\includegraphics[angle=0,height=5in]{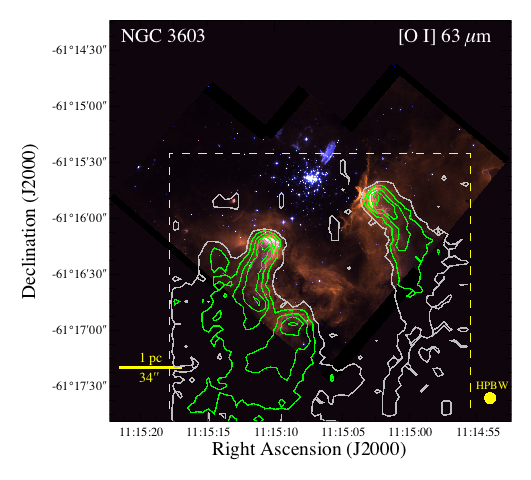} 
\end{center}
\caption{The figure shows the (background) HST false color image (F656N - Red, average of F656N and F658N - Green, and F658N Blue from Brandner et al., 2000) of a 5x4 pc$^{2}$ region around the NGC3603 star cluster.  Discovered almost 200 years ago by John Herschel as a remarkable nebulosity in the constellation Carina, only studies in the last decades have revealed NGC3603’s hidden nature as the Milky Way’s most active stellar nursery. This densely packed condensation of young stars, with an equivalent mass of a few 10000 suns in a volume with a diameter of only 3 light-years across, contains some of the most massive stars known. Their violent winds and ultraviolet radiation have cleared a cavity in the surrounding cloud that now allows unobscured views of the cluster. Deeply embedded in the extended cloud of gas and dust that gave birth to the visible cluster some 1 million years ago, the next generations of stars are forming already. Still invisible in the optical, these sites of ongoing star formation shine bright at far-infrared wavelengths and were the subject of study for GREAT during SOFIAs southern deployments (NGC3603 is visible only from southern skies). The contours correspond to the velocity-integrated intensity maps of the [O I] 63 micron fine-structure line obtained with the HFA, with a half-power beam width of 6.3”. The region of the GREAT map matching the HST image is indicated with dashed lines. The peak intensity is 348 K km/s (with contours in steps of 70 K km/s from the peak). The picture was kindly provided by J.-P. Perez-Beaupuits, priv. comm.}
\label{HFA_map}
\end{figure}

Since mid-2017 GREAT is operated in two configurations only: upGREAT/LFA with HFA and upGREAT/HFA with 4GREAT (Duran et al. in prep), thereby simplifying the operation and minimizing the requests for configuration change, without compromising on the frequency coverage.
Since first light on April 01 2011, GREAT has logged 137 science flights with more than 1000 successful science hours - including 63 flights with upGREAT/LFA, 39 with upGREAT/HFA (25 flights in their joint configuration, since June 2017). More than 200 science projects have been successfully supported. The spatial multiplexing of the arrays now makes it possible to efficiently address large-scale mapping projects within a reasonable investment of observing time. In Fig. \ref{HFA_map} we present an example of a large-scale [O I] 63 ${\mu}m$ fully-sampled map towards the prominent NGC3603 star forming region, covering an area of 160” x 150” (4.7 x 4.4 pc$^{2}$). The map is composed of 2300 spectra, acquired in chopped on-the-fly mode (see \citet{risacher2016b} for the description of the GREAT observing modes) during a short 45 min observing leg in summer 2017 (with excellent atmospheric conditions at 43 kft flight altitude). The sensitivity across the map is very uniform, the noise rms of the data cube (in the velocity channels free of emission) is ~2 K (with a spectral resolution of 0.5 km/s)

Several legacy-type projects have been completed/launched with upGREAT during the last 2 years: e.g. a ~1 square degree region of the Orion Molecular Cloud (Pabst et al. submitted), a fully-sampled map of the M51 galaxy (PIs: J.Pineda, J.Stutzki) is nearing completion. This year the distribution of [CII] and [OI] across the Central Molecular Zone (1.5 x 0.33 deg$^{2}$, composed of 2.3 million spectra) (PIs: A. Harris, R. G{\"u}sten) has been measured during the southern deployment. None of these projects would have been possible without the enhanced mapping speed of upGREAT. All of these examples, together with future legacy projects that will be selected from the explicit calls for legacy proposals starting with observing cycle 7, will contribute to the GREAT legacy of the SOFIA project.

\section{Future work }

The main current limitation for the upGREAT/LFA is the local oscillator system. The solid state multiplier chains provide just enough output power, but are narrow in frequency range and quite fragile. A QCL option would provide much more output power but they are even more narrow band, and having a frequency locked system is not trivial. When the technology allows covering much more bandwidth, with enough output power, the entire range of the 1.8-2.5 THz could be covered at once.

The next natural step for the upGREAT/HFA array would be to populate its second polarization with 7 additional pixels (currently on hold due to lack of funding). The cryostat is already prepared for it and has all the wiring in place. What is needed therefore is to fabricate 7 new mixer blocks with the feedhorns. The LOs have enough power to pump 14 pixels.  Currently, less than 5\% of the LO power is coupled to the mixers, therefore 95\% of it is being terminated into an absorber. It will make sense to use this 95\% of LO power for the second polarization.  
Another future improvement will be to frequency-lock, or better yet, phase-lock the local oscillator, in order to  achieve a better frequency accuracy.

\section*{Acknowledgments}
We thank the SOFIA engineering and operations teams whose support has been essential for the successful installation and operation of the upGREAT instrument. We thank the SOFIA project office at DLR, and in particular A. Himmes and  D. Lilienthal for their outstanding continuous support during the development years of GREAT. 
We are grateful to J.-P. Perez-Beaupuits for kindly providing upGREAT results (Fig. 15) prior to publication.
The development of upGREAT was financed by the participating institutes, by the Federal Ministry of Economics and Technology via the German Space Agency (DLR) under Grants 50 OK 1102, 50 OK 1103 and 50 OK 1104 and within the Collaborative Research Centre 956, sub-projects D2 and D3, funded by the Deutsche Forschungsgemeinschaft (DFG).


\end{document}